\documentclass[nofootinbib,prd,showpacs]{revtex4}
\usepackage{amsmath}
\usepackage{subfigure}
\usepackage{amsfonts}
\usepackage{hyperref}
\usepackage{array}
\usepackage{amssymb}
\usepackage{graphicx}
\graphicspath{ {figs/} }

\usepackage[latin1]{inputenc}
\usepackage[english]{babel}
\usepackage{color,xcolor}
\usepackage{booktabs}
\usepackage{subeqnarray}
\usepackage{url}
\usepackage{paralist}

\usepackage[normalem]{ulem}
\usepackage{times}
\usepackage{makecell}
\usepackage{chngcntr}

\definecolor{purple}{rgb}{1,0,1}
\definecolor{lime}{HTML}{A6CE39} % needs xcolor

\begin{document}
%========================================================
\title{
Black-bounces with multiple throats and anti-throats
}
%=================================================================
%=================================================================
\author{Manuel E. Rodrigues}
\email{esialg@gmail.com}
\affiliation{Faculdade de Ci\^{e}ncias Exatas e Tecnologia, 
Universidade Federal do Par\'{a}\\
Campus Universit\'{a}rio de Abaetetuba, 68440-000, Abaetetuba, Par\'{a}, 
Brazil}
\affiliation{Faculdade de F\'{\i}sica, Programa de P\'{o}s-Gradua\c{c}\~ao em 
F\'isica, Universidade Federal do 
 Par\'{a}, 66075-110, Bel\'{e}m, Par\'{a}, Brazil}
%========================================================
\author{Marcos V. de S. Silva}
\email{marco2s303@gmail.com}
\affiliation{Faculdade de F\'{\i}sica, Programa de P\'{o}s-Gradua\c{c}\~ao em 
F\'isica, Universidade Federal do 
 Par\'{a}, 66075-110, Bel\'{e}m, Par\'{a}, Brazil}
%=================================================================
%-----------------------------------------------------------------
\date{today; \LaTeX-ed \today}
%========================================================
\begin{abstract}
%========================================================
In this article, we test the possibility of building new black-bounce spacetimes with changes in the area in the context of general relativity in four dimensions. These new metrics can present extra structures, such as anti-throats. We see two regions where the area goes to infinity, and, depending on the model, one of these regions presents curvature singularities. Regular metrics can be extended to negative values of the radial coordinate and have a symmetrical structure, whereas some singular cases cannot be extended and have no symmetrical structure. The energy conditions are analyzed, and, for some cases, at least one of the inequalities in the null energy condition is not respected. We also presented models that satisfied the null energy condition outside the event horizon. The event horizon hides the throats of these models.
\bigskip

\end{abstract}
%========================================================
%=================================================================
\maketitle
%=================================================================
\def\HMS{{\scriptscriptstyle{HMS}}}
%========================================================
%\bigskip
%\hrule
%\tableofcontents
%\bigskip
%\hrule
%========================================================
%\parindent0pt
%\parskip7pt
%========================================================
\section{Introduction}
\label{S:intro}
%========================================================
%\bigskip
In the study of black holes, one of the main themes that arise is the singularity problem \cite{Stoica:2014tpa, Ong:2020xwv}. Singularities are points or sets of points where geodesics are interrupted \cite{Bronnikov:2012wsj}, it means that the spacetime time is geodesically incomplete. Since light propagates through null geodesics, the presence of singularities becomes a problem for information propagation \cite{wal}. There are several types of singularities \cite{Tipler:1978zz}. Among them, one of the most commented is the curvature singularities, divergences that appear in the curvature invariants \cite{Bronnikov:2012wsj, Lobo:2020ffi}. The fact that there are no divergences in the curvature invariants tells us only that such a space-time does not present curvature singularities and, therefore, can still present other singularities \cite{Rendall:2005nf}.

Although singularities are expected in black hole solutions, what characterizes a black hole is not the existence of a singularity, but the presence of an event horizon \cite{din,wal}. So, in principle, it is possible to build solutions for black holes that do not present singularities, a regular black hole\footnote{For a review on this topic, see the ref. \cite{Ansoldi:2008jw}.}. Bardeen presented the first nonsingular black hole solution in 1968 \cite{Bardeen}. Despite representing a black hole, the Bardeen solution presented a new parameter that did not allow us to interpret the metric as a solution to Einstein equations for the vacuum. Many years after Bardeen's proposal, Eloy Ayon-Beato and Alberto Garcia considered this parameter to be the charge of a magnetic monopole. They found the Bardeen solution using Einstein equations with a stress-energy tensor for nonlinear electrodynamics \cite{Beato1}. There are various works about regular black holes considering different gravitational theories or nonlinear electrodynamics \cite{Zaslavskii, Fan-Wang, Bronnikov:2017tnz, Capozziello:2014bqa, Rodrigues:2015, NED2, NED3, NED4, NED5, Bronnikov:2006fu, Hollenstein:2008hp, NED10, Rodrigues:2016, Rodrigues:2017, Rodrigues:2018, bambi, neves, toshmatov, DYM, ramon, berej, Rodrigues:2019, Silva:2018, Junior:2020}.

Simpson and Visser proposed another type of regular solution \cite{Simpson:2018tsi}. Like Bardeen's case, this solution has a parameter with no physical interpretation. This parameter modifies the structure of the solution in such a way that there is an area with a nonzero minimum \cite{Simpson:2021vxo}. This minimum represents a throat \cite{Simpson:2018tsi}. Depending on the parameter's value, we can have a black hole, a regular black hole, or a wormhole \cite{Simpson:2018tsi}. This characteristic of having a throat inside the event horizon and the possibility of the solution transiting between a regular black hole and a wormhole is what defines a black-bounce. Recently, new works have emerged to understand the content of matter that forms this type of solution. In these works, the Simpson-Visser element is obtained when considering the presence of nonlinear electrodynamics with a phantom scalar field \cite{Bronnikov:2021uta, Canate:2022gpy}. Other works analyzed the solution properties such as geodesics, absorption, shadows, quasi-normal modes, and others \cite{Simpson:2019cer, Lobo:2020kxn, Tsukamoto:2021caq, Olmo:2021piq, Guerrero:2022qkh, Yang:2021cvh, Lima:2020auu, Lima:2021las}.

In addition to the Simpson-Visser solution, there are other black-bounce solutions. In \cite{Huang:2019arj}, Huang and Yang find a black-bounce solution using a phantom scalar field coupled with linear electrodynamics. In \cite{Lobo:2020ffi}, Lobo and collaborators proposed several black-bounce models. Some of these novel black-bounce solutions present interesting characteristics as more than one photon orbit \cite{Tsukamoto:2021caq}. Extra photon rings generate modifications in the black-bounce shadow \cite{Olmo:2021piq, Guerrero:2022qkh}. These solutions also present gravitational waves echos \cite{Yang:2021cvh}.

The Simpson-Visser solution and the novel black-bounce solutions can represent a wormhole if the new parameter takes on specific values. These wormholes have only one throat. In the literature, due to the couple with a scalar field, there are wormhole models with more than one throat, and even anti-throats \cite{Guendelman:2015wsv, Hoffmann:2017vkf, Hoffmann:2018oml, Bakopoulos:2018nui, Ibadov:2020ajr, Ibadov:2020btp}, which, considering a spherically symmetric spacetime, is a $2$-dimensional constant-time hyper-surface of local maximum area \cite{Anabalon:2018rzq}. So, it may be possible to build solutions that present a wormhole with more than one throat/anti-throat at some limit. This type of solution will require modifications to the wormhole area since the presence of the throat/anti-throat arises from the structure of the wormhole area. In this way, since the black-bounce models presented so far can only interpolate between black holes and wormholes with a single throat, it is interesting to seek new proposals that can encompass other types of wormholes. Therefore, new black-bounce models should be able to, depending on the chosen parameters, describe wormholes with multiple throats/anti-throats. This is the type of model we aim to explore through this work.

The structure of this article is organized as follows. In SEC. \ref{S:general}, we present in general the tools that will be used to analyze the metrics, such as regularity and energy conditions. The new singular and regular black-bounce are present in SEC. \ref{S:new}. Section \ref{S:conclusion} presents our conclusions and perspectives.

 We adopt the metric signature $(+,-,-,-)$. We shall work in geometrodynamic units where $G=\hbar=c=1$.

%========================================================
\section{General black-bounce spacetimes}
\label{S:general}
%========================================================
\subsection{Metric and curvature}\label{SS:metric}
The most general static spherically symmetric metric locally can always be cast into the form:
\begin{equation}
ds^2 = f(r) dt^2 - \frac{dr^2}{f(r)}  - 
\Sigma(r)^2 (d \theta^2 +\sin^2 \theta d \phi^2).\label{ele}
\end{equation}
The area of a sphere at radial coordinate $r$ is now $A(r) = 4\pi \Sigma(r)^2$, and we shall use the freedom to choose $\Sigma(r)$ extensively in the discussion below.

In order to investigate the regularity of the spacetime, we may also calculate the Kretschmann scalar, $K=R_{\alpha\beta\mu\nu}R^{\alpha\beta\mu\nu}$, in terms of the Riemann components, which is written as~\cite{Bronnikov:2012wsj,Lobo:2020ffi}
\begin{eqnarray}
&&K=\frac{
(\Sigma^2 f'')^2+2(\Sigma f' \Sigma')^2 +2\Sigma^2(f' \Sigma'+2f \Sigma'')^2
+4(1-f\Sigma'^2)^2}
{\Sigma^4}\,.
\label{Kret3}
\end{eqnarray}
For the spacetime be asymptotically flat, the functions must satisfy the condition $\lim_{r\to\infty} \{f(r),\Sigma(r)/r\}=\{1,1\}$. In this limit, the Kretschmann scalar is null.
%==========================================
\subsection{Energy conditions}\label{SS:stress-energy}
%==========================================
The Einstein equations are given by \cite{wal}
\begin{eqnarray}
R_{\mu\nu}-\frac{1}{2}g_{\mu\nu}{R}=\kappa^2 T_{\mu\nu}\,,\label{eqEinstein}
\end{eqnarray}
where $g_{\mu\nu}$ is the metric tensor, $R_{\mu\nu}=R^{\alpha}{}_{\mu\alpha\nu}$, ${R}=g^{\mu\nu}R_{\mu\nu}$, $T_{\mu\nu}$ the stress-energy tensor and $\kappa^2=8\pi$. If we consider the matter sector as an anisotropic fluid, then the mixed components of the stress-energy tensor are given by \cite{din}
\begin{eqnarray}
T^{\mu}{}_{\nu}={\rm diag}\left[\rho,-p_r,-p_t,-p_t\right]\,,\label{EMT}
\end{eqnarray}
where $\rho$, $p_r$, and  $p_t$ are the energy density, radial pressure and tangential pressure, respectively. Taking into account the line element \eqref{ele}, the Einstein equations \eqref{eqEinstein} provide the following stress-energy profile
\begin{eqnarray}
&&\rho=-\frac{\Sigma \left(f' \Sigma'+2 f \Sigma''\right)+f \Sigma'^2-1}{\kappa ^2 \Sigma^2}\label{density}\,,\\
&&p_r=\frac{\Sigma f'\Sigma'+f \Sigma'^2-1}{\kappa ^2 \Sigma^2}\,,\label{pr}\\
&&p_t=\frac{\Sigma f''+2 f' \Sigma'+2 f \Sigma''}{2 \kappa ^2 \Sigma}\label{pt}\,.
\end{eqnarray}

To situations where $f(r)<0$, $r$ and $t$ are the timelike and spacelike coordinates, respectively, we must have
\begin{eqnarray}
T^{\mu}{}_{\nu}={\rm diag}\left[-p_r,\rho,-p_t,-p_t\right]\,,\label{EMT2}
\end{eqnarray}
The fluid quantities, in this region, are
\begin{eqnarray}
&&\rho=-\frac{\Sigma f'\Sigma'+f \Sigma'^2-1}{\kappa ^2 \Sigma^2}\label{density2}\,,\\
&&p_r=\frac{\Sigma \left(f' \Sigma'+2 f \Sigma''\right)+f \Sigma'^2-1}{\kappa ^2 \Sigma^2}\,,\label{pr2}\\
&&p_t=\frac{\Sigma f''+2 f' \Sigma'+2 f \Sigma''}{2 \kappa ^2 \Sigma}\label{pt2}\,.
\end{eqnarray}

At the any possible event horizon, $f(r)=0$, we find
\begin{equation}
    T^0_{\ 0}=T^1_{\ 1}=-\frac{\Sigma f'\Sigma'-1}{\kappa^2\Sigma^2}, \qquad T^2_{\ 2}=T^3_{\ 3}=\frac{\Sigma f''+2f'\Sigma'}{2\kappa^2\Sigma^2}.
\end{equation}
This result tell us that the energy density and the radial pressure are continuous in any possible horizon.

The standard energy conditions~\cite{book} for the stress-energy tensor \eqref{EMT}, are given by the inequalities\footnote{Subindices 1 and 2 appear when the energy condition is related to radial and tangential pressure, respectively. In the case of subindice 3, there are two different situations; the first is when the energy condition is related only to the energy density, which is the dominant and weak cases; the second is when we have the presence of all fluid quantities, which is the strong case.}
\begin{eqnarray}
&&NEC_{1,2}=WEC_{1,2}=SEC_{1,2} 
\Longleftrightarrow \rho+p_{r,t}\geq 0,\label{Econd1} \\
&&SEC_3 \Longleftrightarrow\rho+p_r+2p_t\geq 0,\label{Econd2}\\
&&DEC_{1,2} \Longleftrightarrow \rho-|p_{r,t}|\geq 0 \Longleftrightarrow 
(\rho+p_{r,t}\geq 0) \hbox{ and } (\rho-p_{r,t}\geq 0),\label{Econd3}\\
&&DEC_3=WEC_3 \Longleftrightarrow\rho\geq 0,\label{Econd4}
\end{eqnarray}
As part of the dominant energy condition is already in the null energy condition, we will consider only $DEC_{1,2}\Longrightarrow \rho-p_{r,t}\geq 0$.

Outside any event horizon that might be present we must have $f(r)>0$. Also $\Sigma(r)>0$ everywhere.
So we easily verify that $NEC_1=WEC_1=SEC_1$ all exhibit negative values outside the event horizon whenever $\Sigma''(r)>0$.  Thus the NEC, and so all energy conditions, are violated for black-bounce models whenever $\Sigma''(r)>0$.

As $\rho$ and $p_r$ are continuous across the horizon, $NEC_1=WEC_1=SEC_1$ and $DEC_1$ are also continuous. As the energy density is continuous, $WEC_3$ also is. The remain energy conditions are not necessarily continuous across the event horizon.

%==========================================
\section{New black-bounce spacetimes}\label{S:new}
%==========================================
The Simpson-Visser model to black-bounce is given by
\begin{equation}
\Sigma(r)=\sqrt{r^2+a^2},\qquad f(r)=1-\frac{2m}{\sqrt{r^2+a^2}}\,.\label{Visser}
\end{equation}
To find new black-bounces that generalize the Simpson--Visser model, we will fix $f(r)$ as \eqref{Visser} and test different models of $\Sigma(r)$. We first consider a general black-bounce model that is given by
\begin{equation}
\Sigma(r)=\sqrt{\left(r^{2n_1}+d^{2n_1}\right)^{1/n_1}e^{\left(b^2/(c_1r^2+c_2r+c_3)\right)^{n_2}}},\qquad f(r)=1-\frac{2m}{\sqrt{r^2+a^2}}\,.\label{newBBS}
\end{equation}
Constants $n_1$ and $n_2$ are integers and non-negative numbers, while the constants $a$, $b$, $d$, and $c_i$ have no restrictions on being integers or fractions, but must always be non-negative. The constants $b$, $d$, and $c_2$ have the dimension of length, $c_3$ length squared, and $n_1$, $n_2$, and $c_1$ are dimensionless. The form of the function $\Sigma(r)$ is chosen such that we can recover the Simpson--Visser solution by an appropriate choice of parameters, and that we can recover Schwarzschild and Minkowski solutions in the appropriate limit. To $d=a$, $n_1=1$ and $b=0$ we recover the Simpson--Visser solution. Furthermore, the shape of the $\Sigma(r)$ allows us to obtain models with more complex causal structures, including, for example, multiple throats and the addition of anti-throats. Additionally, it enables us to locally satisfy energy conditions that were previously always violated in known black-bounce spacetimes.
\subsection{Singular spacetimes}
Here are some models, whose Kretschmann scalar presents divergences.
\subsubsection{Model $d=0, n_1=n_2=c_1=1, c_2=0$ and $c_3=0$}
For the case $d=0$, $n_1=n_2=c_1=1$, $c_2=0$ and $c_3=0$ in \eqref{newBBS}, we have
\begin{equation}
\Sigma(r)=\sqrt{r^2e^{b^2/r^2}},\qquad f(r)=1-\frac{2m}{\sqrt{r^2+a^2}}\,.\label{fnewBBS}
\end{equation}
The area becomes
\begin{equation}
A=4\pi r^2e^{b^2/r^2}.
\end{equation}

This area is symmetric to $r\rightarrow -r$. In the Simpson--Visser solution, the area is infinite to $r\rightarrow \pm \infty$. To the model \eqref{fnewBBS}, there is no need to consider negative radial coordinate values, since the area is infinity to $r\rightarrow\infty$ and $r\rightarrow 0$. To analyze the maximums and minimums we need to solve $A'(r)=0$. We find that the area has extremes values at $r=\pm b$, where $A=4\pi b^2 e$ and $A''=16\pi e>0$. As $A''(r=\pm b)>0$, $r=\pm b$ represents minimum values of the area. In this case, $b$ is a throat. Let us impose $r_T=\sqrt{e} b<r_H=\sqrt{4m^2-a^2}$, where $r_T$ is the radius of the throat and $r_H$ is the radius of the event horizon, so that the throat is located inside the event horizon.

As the Simpson--Visser model, $f(r)$, $f(r)'$, and $f(r)''$, to the model \eqref{fnewBBS}, are finite everywhere, in other words, these functions satisfy the condition of regularity. The function $\Sigma$ is non-zero everywhere, so also satisfies the condition of regularity. The problem arises due to the fact that $\Sigma'(r)$ and $\Sigma''(r)$ are singular at $r=0$. So that, we are not able to guarantee the regularity of the spacetime and we need to analyze $K$. At $r=0$ the term that dominates is
\begin{eqnarray}
    K\approx 
  \frac{12 b^8 (a-2
   m)^2}{a^2 r^{12}} + \frac{24 b^8 m (a-2 m)}{a^4 r^{10}}+O\left(\frac{1}{r^8}\right).
\end{eqnarray}
There is a curvature singularity at $r=0$. If $a=2m$, the Kretschmann scalar behaves as $K\propto r^{-8}$ to $r\rightarrow 0$. Thus, this spacetime is composed of an event horizon which contains a singularity at $r=0$, and a throat located between the event horizon and the singularity. The causal structure of this spacetime and the others presented in this study can be analyzed through the Carter-Penrose diagrams, which are provided in Appendix \ref{Appen_A}.

From the null energy condition, we find that
\begin{eqnarray}
&&NEC_1\Longleftrightarrow
-\frac{2 b^2 \left(b^2+r^2\right) }{\kappa ^2 r^6} \left(1-\frac{2m}{\sqrt{a^2+r^2}}\right)
\geq 0 \label{NEC1fnewBBS},\quad \mbox{to} \quad f(r)>0,\\
&&NEC_1\Longleftrightarrow
\frac{2 b^2 \left(b^2+r^2\right) }{\kappa ^2 r^6}\left(1-\frac{2m}{\sqrt{a^2+r^2}}\right)
\geq 0,\quad \mbox{to} \quad f(r)<0.\label{NEC11fnewBBS}
\end{eqnarray}
It is known that, if the null energy condition is violated, all energy conditions are also violated. From Eq. \eqref{NEC1fnewBBS}, $NEC_1$ will be satisfied only to $r\leq \sqrt{4m^2-a^2}$. However, this region is inside the event horizon, where $t$ is the spacelike coordinate, and Eq. \eqref{NEC1fnewBBS} has no validity. Inside the event horizon, $NEC_1$ is determined by Eq. \eqref{NEC11fnewBBS} and, according to this equation, is violated in that region. By properly choosing the parameters, it is possible to satisfy $NEC_2$ outside the horizon, however, as $NEC_1$ is not satisfied, the null energy condition is still violated.

\subsubsection{Model $d=0$, $n_1=n_2=1$, $c_1=0$ and $c_2=m$}
For the case $d=0$, $n_1=n_2=1$, $c_1=0$ and $c_2=m$  in \eqref{newBBS}, we have
\begin{equation}
\Sigma(r)=\sqrt{r^2 e^{\frac{b^2}{m r+c_3}}},\qquad f(r)=1-\frac{2m}{\sqrt{r^2+a^2}}\,.\label{snewBBS}
\end{equation}
The area becomes
\begin{equation}
A=4\pi r^2e^{\frac{b^2}{m r+c_3}}.
\end{equation}
This area is not symmetric to $r\rightarrow -r$. Solving $dA/dr=0$, we find 
\begin{equation}
r\rightarrow r_1= 0, \qquad r\rightarrow r_2=\frac{b^2 m-b m \sqrt{b^2-8 c_3}-4 m c_3}{4 m^2}, \qquad r\rightarrow r_3=\frac{b^2 m+b m \sqrt{b^2-8 c_3}-4 m c_3}{4 m^2}.\label{extremaModel2}
\end{equation}
We need $b^2/8>c_3>0$ to $r_2$ and $r_3$ be real and different from each other. If $c_3=0$, we find that
the area has only one extreme value, which is $r=b^2/(2 m)$. This point is a minimum, once we have $A''(r=b^2/(2m))=8\pi e^2>0$. This structure is similar to the previous case with the difference that the throat is now located in $r=b^2/(2 m)$.

In the limit $r\rightarrow 0$, we find that the Kretschmann scalar, with $c_3=0$, behaves as
\begin{equation}
    K\approx \frac{3 b^8 (a-2 m)^2}{4 a^2 m^4 r^8}-\frac{2b^6 (a-2 m)^2}{a^2 m^3r^7 }+O\left(\frac{1}{r^6}\right),
\end{equation}
so that, there is a curvature singularity in $r=0$. For the case $a=2m$, the Kretschmann scalar diverges at $r\rightarrow 0$ as $K\propto r^{-4}$.

From the null energy condition, we find
\begin{eqnarray}
&&NEC_1\Longleftrightarrow
-\frac{b^4 \left(1-\frac{2 m}{\sqrt{a^2+r^2}}\right)}{2 \kappa ^2 m^2 r^4}
\geq 0,\quad \mbox{to} \quad f(r)>0,\\
&&NEC_1\Longleftrightarrow
\frac{b^4 \left(1-\frac{2 m}{\sqrt{a^2+r^2}}\right)}{2 \kappa ^2 m^2 r^4}
\geq 0,\quad \mbox{to} \quad f(r)<0.
\end{eqnarray}
In addition to $NEC_1$, $NEC_2$ is also violated both inside and outside the event horizon. We see that the null energy condition is everywhere violated.

In the case where $c_3=b^2/8$, we get
\begin{eqnarray}
    A=4 \pi  r^2 e^{\frac{8 b^2}{b^2+8 m r}},\quad 
    A'=\frac{8 \pi  r e^{\frac{8 b^2}{b^2+8 m r}} \left(b^2-8 m r\right)^2}{\left(b^2+8 m r\right)^2}.
\end{eqnarray}
The area now tends to infinity to $r\rightarrow -b^2/8m$ and not to $r\rightarrow 0$ as before. Now there are two extrema points, $r=0$ and $r=b^2/8m$. In the extrema points we find $A''(r=0)=8 e^8 \pi$ and $A''(r=b^2/8m)=0$, which means that we have a minimum point and an inflection point.

\begin{figure}
    \centering
    \includegraphics[scale=0.7]{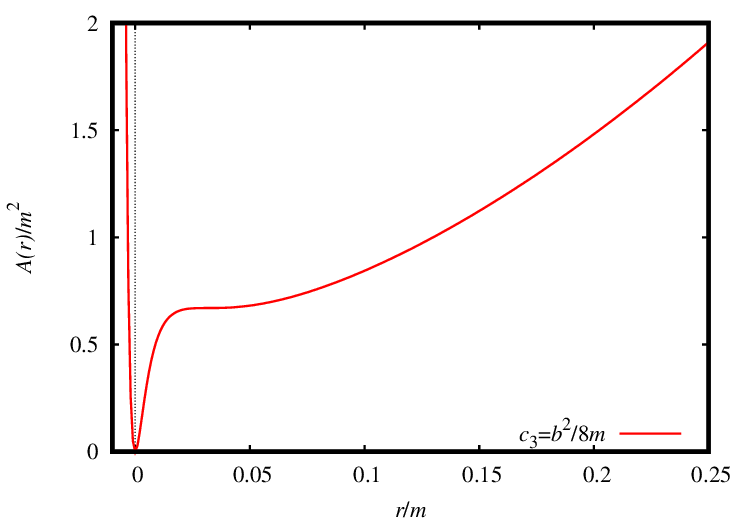}
    \includegraphics[scale=0.7]{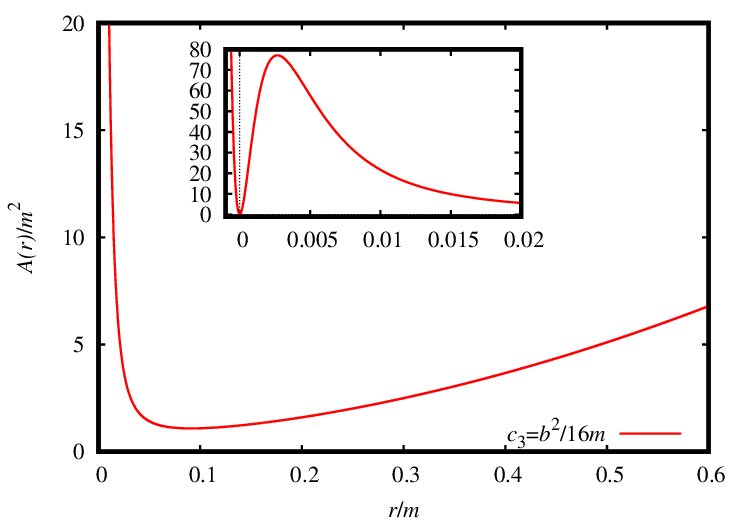}
    \includegraphics[scale=0.7]{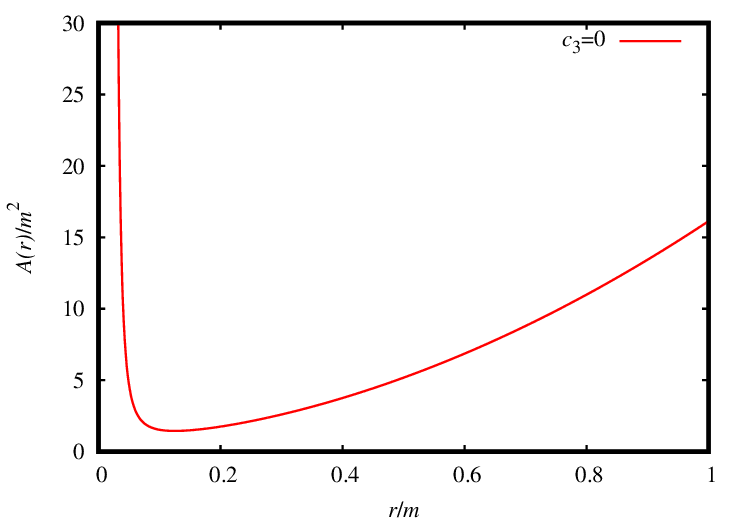}
    \caption{Graphic representation of the area associated to the model \eqref{snewBBS} with $b=0.5m$.}
    \label{fig:areamodel2}
\end{figure}

In general, there are three extrema points, \eqref{extremaModel2}. In Fig. \ref{fig:areamodel2} we see that to $c_3=0$ there is only a nonzero minimum point, which represents a throat. To $c_3=b^2/16m$ there are two minimum, but one of them is at $r=0$, and a maximum, which is an anti-throat. To $c_3=b^2/8m$ there is a minimum and a inflection point. If $c_3\neq0$, $r\rightarrow 0$ is point of minimum.

The analytical expression for the Kretschmann scalar is not simple but to $r\rightarrow 0$ we find
\begin{equation}
    K \approx \frac{4 e^{-\frac{2 b^2}{c_3}} \left(a-(a-2 m) e^{\frac{b^2}{c_3}}\right)^2}{a^2
   r^4}+O\left(\frac{1}{r^3}\right).
\end{equation}
Even with $c_3\neq 0$, the metric presents a divergence to $r\rightarrow 0$. So that, to $c_3\neq 0$, there is only one region that the area goes to infinite, $r\rightarrow \infty$. Even for $a=2m$, we find $K\propto r^{-4}$ at $r\rightarrow 0$. As $r=0$ is a singularity, one of the throats becomes singular. Thus, only one of them is traversable.

\begin{figure}
    \centering
    \includegraphics[scale=0.7]{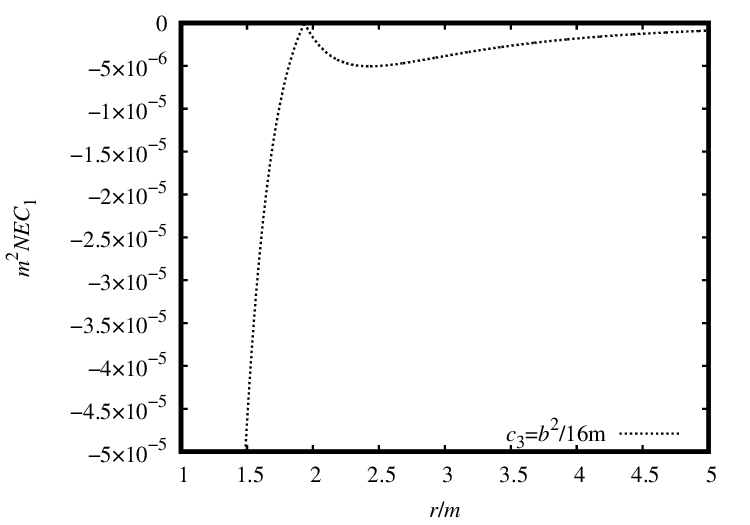}
    \caption{Condition $NEC1$ related to the model \eqref{snewBBS} with $a=0.5m$, $c_3=b^2/16m$, and $b=0.5m$.}
    \label{fig:ECModel2}
\end{figure}

In Fig. \ref{fig:ECModel2} we see that the condition $NEC_1$ is everywhere violated. Even for $c_3\neq 0$, $NEC_2$ is not satisfied in the entire region inside or outside the horizon, so the null energy condition is always violated.

\subsubsection{Model $n_1=n_2=1$, $c_1=0$, $b=d$, and $c_2=m$}
\begin{figure}[!htb]
	\includegraphics[scale=0.70]{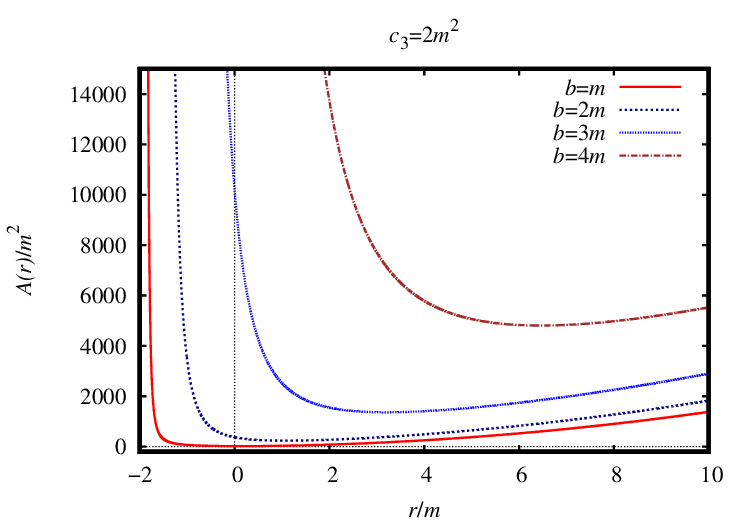}
	\includegraphics[scale=0.7]{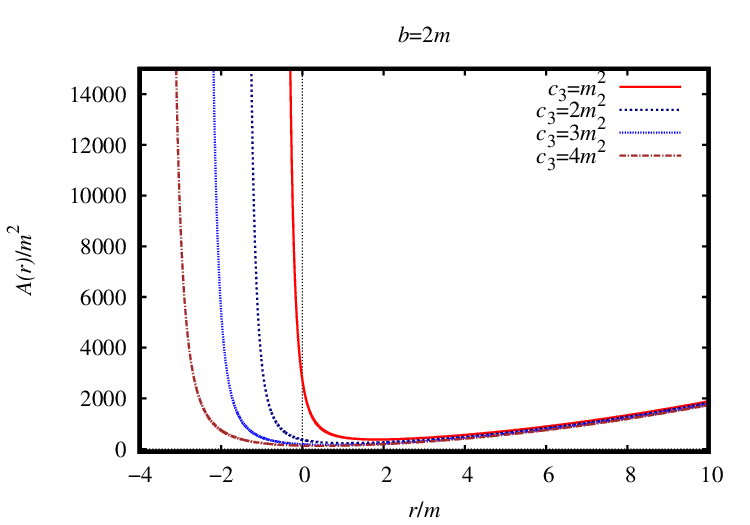}
	\caption{Graphical representation of the area to the model \eqref{tnewBBS}.}\label{figtnewBBS}
\end{figure}

For the case $n_1=n_2=1$, $c_1=0$, $b=d$, and $c_2=m$ in \eqref{newBBS}, we have
\begin{equation}
\Sigma(r)=\sqrt{\left(b^2+r^2\right) e^{\frac{b^2}{m r+c_3}}},\qquad f(r)=1-\frac{2m}{\sqrt{r^2+a^2}}\,.\label{tnewBBS}
\end{equation}
The area becomes
\begin{equation}
A=4\pi\left(b^2+r^2\right) e^{\frac{b^2}{m r+c_3}}.
\end{equation}
This area is not symmetric to $r\rightarrow -r$. The extremes values of $A(r)$ will exist only for some values of $b$, $d$ and $c_3$. To negative values of $c_3$ the area diverges in the positive part of $r$. If $c_3$ is positive we have only one root of $dA(r)/dr$. To $c_3<0$ there are three roots, however, only one of the roots is located before the point where the area diverges. In Fig. \ref{figtnewBBS} we see how behaves the area to this black-bounce. As $c_3$ increases the point where area diverges and point of minimum are dislocated to the left. If $b$ increases the minimum point is also dislocated to the left.

For the Kretschmann scalar, the analytical expression in not simple but we can analyze some limits. To $r\rightarrow \infty$ we find
\begin{equation}
    K\approx \frac{48m^2}{r^6}+O\left(\frac{1}{r^7}\right).
\end{equation}
According to this result the spacetime is regular in the infinity and asymptotically flat. To $r\rightarrow 0$ we find the $K$ tends to a constant. Different from the case before, this metric is regular at $r\rightarrow0$.
To $r\rightarrow -c_3/m$ we get
\begin{eqnarray}
    K\approx\frac{3 b^8 m^4 \left(\sqrt{m^2a^2+c_3^2}-2m^2 \right)^2}{4 \left(a^2 m^2+c_3^2\right) (m r+c_3)^8}
   +O\left(\frac{1}{(mr+c_3)^7}\right).
\end{eqnarray}
There is a singularity at $r\rightarrow -c_3/m$.

\begin{figure}[!htb]
	\includegraphics[scale=0.7]{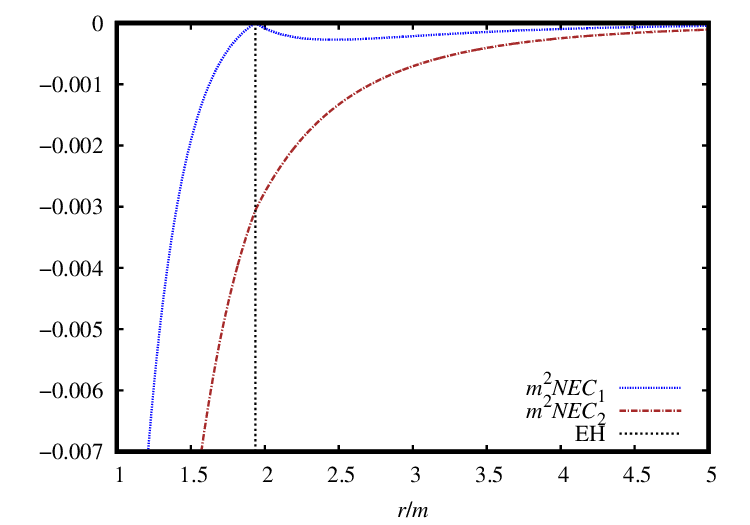}
	\caption{Null energy condition to the model \eqref{tnewBBS} with $b=m$, $c_3=m^2$ and $a=0.5m$.}\label{FigECtnewBBS}
\end{figure}

As the expressions to the null energy condition are not simple, in Fig. \ref{FigECtnewBBS} we analyze them graphically.
The null energy condition is always violated. So, all energy conditions are violated.
\subsubsection{Model $\Sigma(r)=\frac{m}{\sqrt{e^{-\frac{m^2\left(m-8r\right)}{8r^3}}-1}}$}
We will present a slightly different model from the previous ones, given by
\begin{equation}
\Sigma(r)=\frac{m}{\sqrt{e^{-\frac{m^2\left(m-8r\right)}{8r^3}}-1}},\qquad f(r)=1-\frac{2m}{\sqrt{r^2+a^2}}\,.\label{LSnewBBS}
\end{equation}
The area becomes
\begin{equation}
    A=\frac{4 \pi  m^2}{e^{-\frac{m^2 (m-8 r)}{8 r^3}}-1}.
\end{equation}
For large values of the radial coordinate we get
\begin{equation}
    A(r)\approx 4\pi r^2+ O(r),\quad \mbox{to} \quad r\rightarrow \infty.
\end{equation}
At $r\rightarrow 3m/16$ we find $A'(r)=0$ and $A''(r)=1.55076$, which means that there is a throat.

In the limit $r\rightarrow m/8$, the Kretschmann scalar is 
\begin{equation}
    K\approx \frac{19 \left(\sqrt{64 a^2+m^2}-16 m\right)^2}{4 \left(64 a^2+m^2\right) \left(r-\frac{m}{8}\right)^4}+O\left(\frac{1}{ \left(r-\frac{m}{8}\right)^3}\right),
\end{equation}
which means that there is a singularity. For the case $a=\sqrt{255} m/8$, as $r\rightarrow 0$, we find $K\propto (r-m/8)^{-2}$.

\begin{figure}
    \centering
    \includegraphics[scale=0.7]{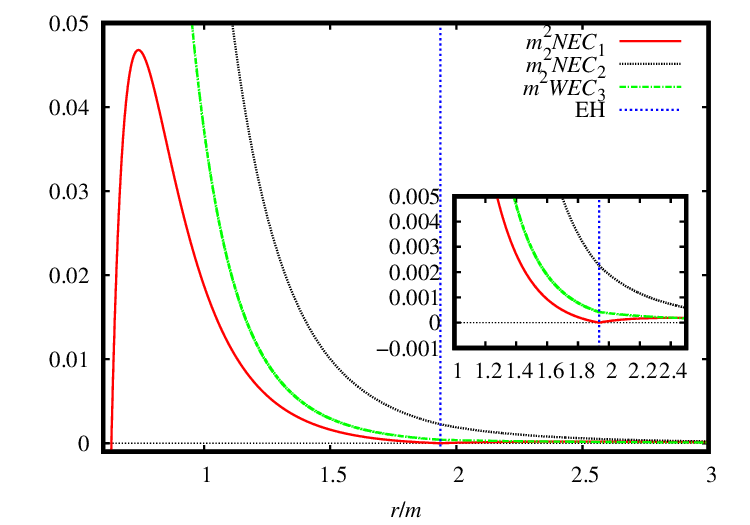}
    \caption{Energy conditions associated with \eqref{LSnewBBS} with $a=0.5m$.}
    \label{fig:modelextrasingular}
\end{figure}

In Fig. \ref{fig:modelextrasingular} we see the energy conditions to the model \eqref{LSnewBBS}. For certain values of $a$, the null energy condition can be satisfied outside the event horizon. The function $WEC_3$ is always positive, so that, the energy density is also positive. 
\subsection{Regular spacetimes}
Here are some models, whose Kretschmann scalar presents no divergences.
\subsubsection{Model $n_1=n_2=c_1=1$, and $c_2=0$}
For the case $n_1=n_2=c_1=1$, and $c_2=0$ in \eqref{newBBS}, we have
\begin{equation}
\Sigma(r)=\sqrt{\left(d^2+r^2\right) e^{\frac{b^2}{c_3+r^2}}},\qquad f(r)=1-\frac{2m}{\sqrt{r^2+a^2}}\,.\label{fRnewBBS}
\end{equation}
The area becomes
\begin{equation}
A=4 \pi  \left(d^2+r^2\right) e^{\frac{b^2}{r^2+c_3}}.
\end{equation}
This area is symmetric to $r\rightarrow -r$ only if $c_3>0$. Solving $dA/dr=0$ we find that the area has extremes values at
\begin{eqnarray}
&&r\rightarrow r_1=0,\quad r\rightarrow r_2=-\frac{\sqrt{-b \sqrt{b^2+4 d^2-4 c_3}+b^2-2 c_3}}{\sqrt{2}},\quad r\rightarrow r_3=\frac{\sqrt{-b \sqrt{b^2+4 d^2-4 c_3}+b^2-2 c_3}}{\sqrt{2}},\nonumber\\
&&r\rightarrow r_4=-\frac{\sqrt{b \sqrt{b^2+4 d^2-4 c_3}+b^2-2 c_3}}{\sqrt{2}},\qquad r\rightarrow r_5=\frac{\sqrt{b \sqrt{b^2+4 d^2-4 c_3}+b^2-2 c_3}}{\sqrt{2}}.
\end{eqnarray}

\begin{figure}[!htb]
	\includegraphics[scale=0.70]{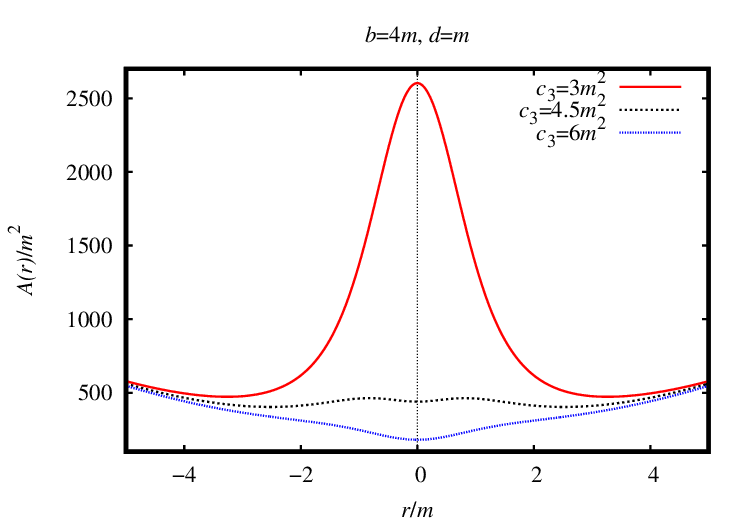}
	\includegraphics[scale=0.7]{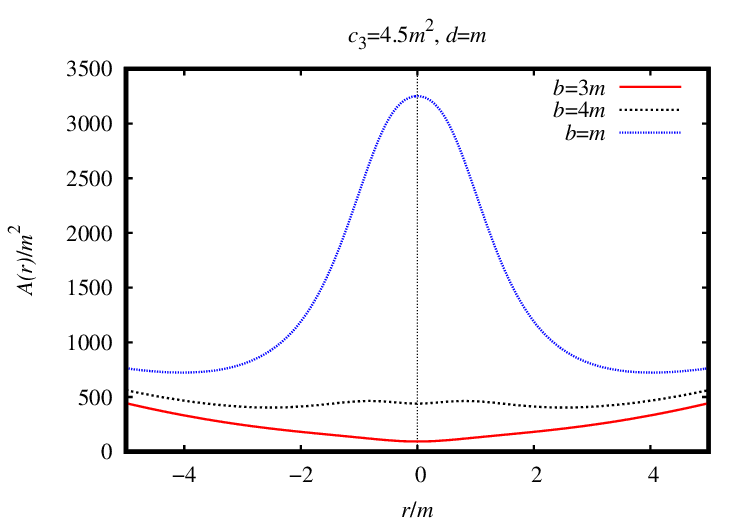}
	\includegraphics[scale=0.7]{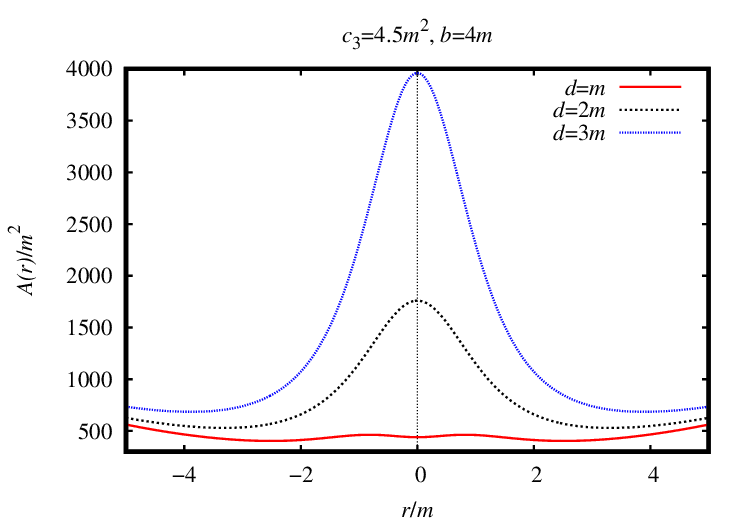}
	\caption{Graphical representation of the area to the model \eqref{fRnewBBS}.}\label{fig1fRnewBBS}
\end{figure}

In Fig. \ref{fig1fRnewBBS}, we represent the area for different values of $c_3$, $b$ and $d$. We see that, depend on the parameters, there is one or more throats. There are also the presence of anti-throats.

For the Kretschmann scalar, we see that we have no curvature singularity once, to $r\rightarrow0$, $K$ tends to a constant, and, 
 to $r\rightarrow \pm \infty$, $K$ behaves as $r^{-6}$.

\begin{figure}[!htb]
	\includegraphics[scale=0.70]{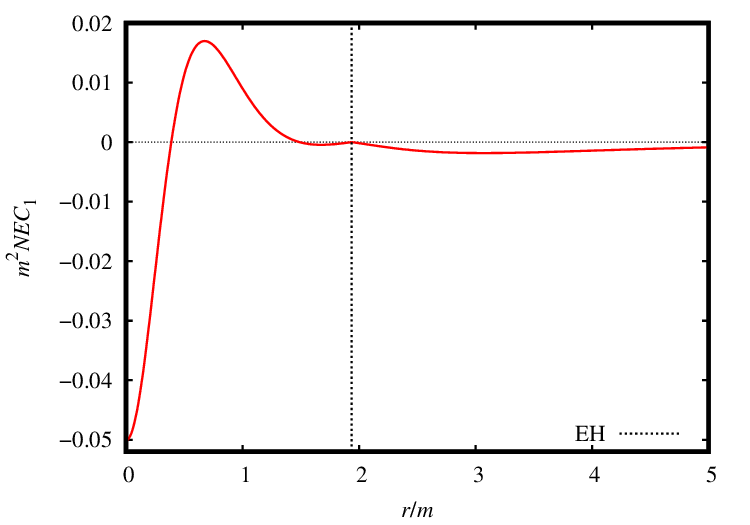}
	\includegraphics[scale=0.70]{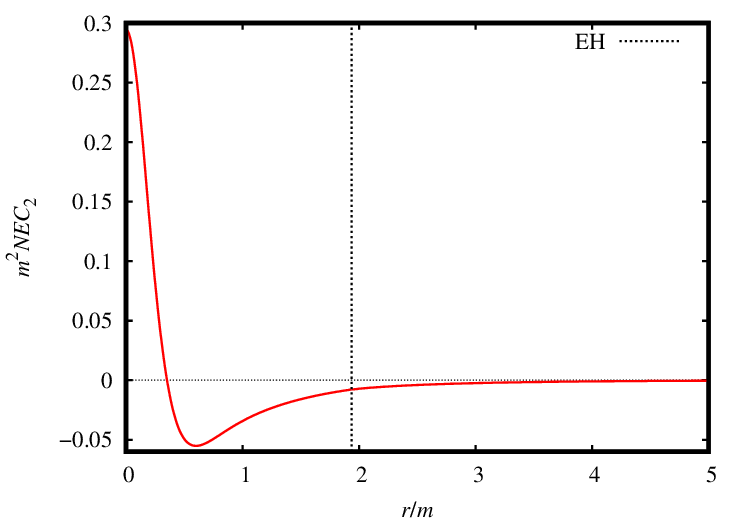}
	\caption{Null energy condition to the model \eqref{fRnewBBS} with with $d=m$, $b=4m$, $c_3=4.5m$ and $a=0.5m$. As the functions are symmetric, we will only consider the positive part of the radial coordinate.}\label{fig3fRnewBBS}
\end{figure}

The null energy condition is analyzed in Fig. \ref{fig3fRnewBBS}. There are always some region of the spacetime where the null energy condition is violated.

\subsubsection{Model $\Sigma(r)=\sqrt{b^2+r^2 \cos \left(\frac{d}{\sqrt{d^2+r^2}}\right)}$}
Let us consider the model
\begin{equation}
\Sigma(r)=\sqrt{b^2+r^2 \cos \left(\frac{d}{\sqrt{d^2+r^2}}\right)},\qquad f(r)=1-\frac{2m}{\sqrt{r^2+a^2}}\,.\label{SRnewBBS}
\end{equation}
The area becomes
\begin{equation}
A=4 \pi  \left(b^2+r^2 \cos \left(\frac{d}{\sqrt{d^2+r^2}}\right)\right).
\end{equation}
This area is symmetric to $r\rightarrow -r$. The area has only one extreme point that is located at $r\rightarrow 0$. In this point we find
\begin{equation}
A(r=0)=4b^2\pi,\quad A'(r=0)=0,\quad \mbox{and} \quad
A''(r=0)=8\pi \cos (1).
\end{equation}
As the second derivative of the area is positive, there is a point of minimum, which represents a throat.

In the limit $r\rightarrow0$ the Kretschmann scalar tends to a constant, and to $r\rightarrow\pm\infty$ $K$ behaves as $r^{-6}$. So the spacetime has no curvature singularities.

\begin{figure}[!htb]
	\includegraphics[scale=0.70]{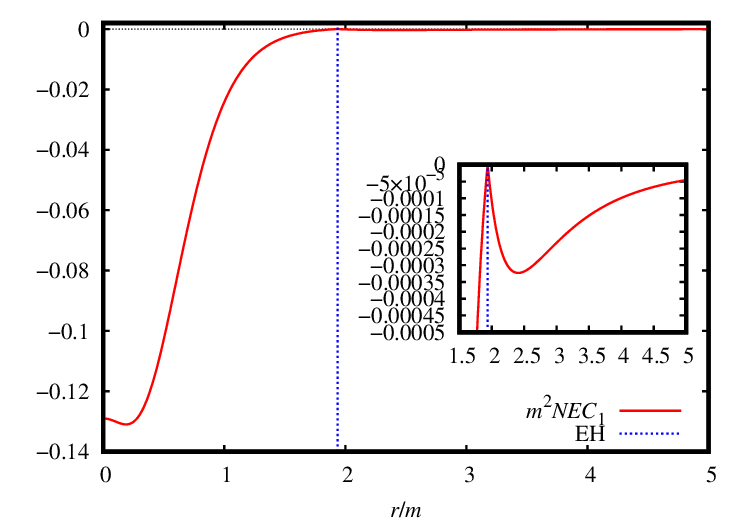}
	\includegraphics[scale=0.70]{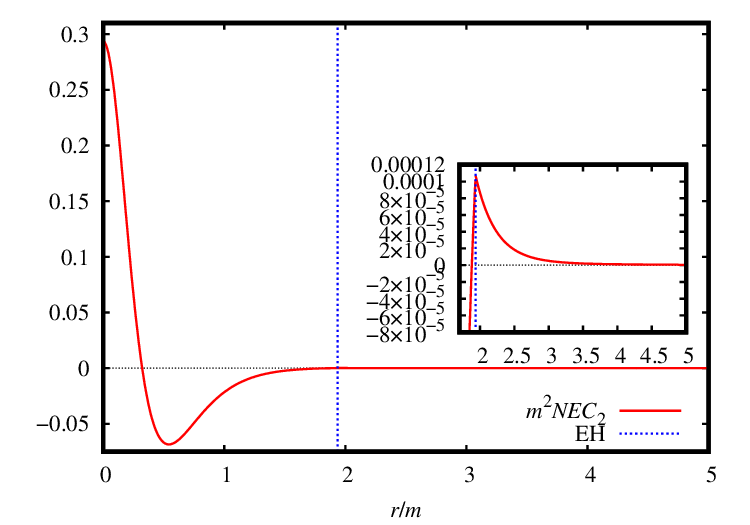}
	\caption{Null energy conditions to the model \eqref{SRnewBBS} with $b=d=2a=m$.}\label{fig2SRnewBBS}
\end{figure}

The null energy condition is analyzed in Fig. \ref{fig2SRnewBBS}. As $NEC_1$ is violated every, the null energy condition is violated.

\subsubsection{Model $\Sigma(r)=\sqrt{b^2+\frac{r^{2n+2}}{d^{2 n}+r^{2 n}}}$}
Let us consider the model
\begin{equation}
\Sigma(r)=\sqrt{b^2+\frac{r^{2n+2}}{d^{2 n}+r^{2 n}}}\,,\label{TRnewBBS}
\end{equation}
with $n$ being a integer number.
The area becomes
\begin{equation}
A=4 \pi\left( b^2+\frac{r^{2 n+2}}{d^{2 n}+r^{2 n}}\right).
\end{equation}
This area is symmetric to $r\rightarrow -r$. The area has only one real extreme value at $r\rightarrow 0$. Actually, to $n$ being a positive and integer number, we have
\begin{equation}
    \left.\frac{d^{2n+1}A}{dr^{2n+1}}\right|_{r=0}=0, \ \mbox{and}\ \left.\frac{d^{2n+2}A}{dr^{2n+2}}\right|_{r=0}>0.
\end{equation}
So this point is a throat.

In the limit $r\rightarrow 0$, there are two possible results to the Kretschmann scalar, depending on the value of $n$, which are
\begin{eqnarray}
K&\approx& \frac{2 \left(1-\frac{2 m}{\sqrt{a^2}}\right)^2}{b^4}+\frac{4 m^2}{a^6}+\frac{4}{b^4}+ O\left(r^2\right), \mbox{to} \ n=0,\\
K &\approx& \frac{4 m^2}{a^6}+\frac{4}{b^4}+ O\left(r^2\right), \mbox{to} \ n>1.
\end{eqnarray}
In the infinity of the radial coordinate we find
\begin{eqnarray}
K&\approx&\frac{4}{r^4}+\frac{16 m}{r^5}+\frac{48 m^2}{r^6}\mp \frac{8 \left(a^2 m+4 b^2 m\right)}{r^7}+O\left(\frac{1}{r^8}\right), \mbox{to} \ n=0\ \mbox{and} \ r\rightarrow \pm \infty,\\
K&\approx& \frac{48 m^2}{r^6}\pm\frac{32 \left(b^2 m-d^2 m\right)}{r^7}+O\left(\frac{1}{r^8}\right), \mbox{to} \ n=1\ \mbox{and} \ r\rightarrow \pm \infty,\\
K&\approx&\frac{48 m^2}{r^6}\pm\frac{32 b^2m}{r^7}+O\left(\frac{1}{r^8}\right), \mbox{to} \ n>1\ \mbox{and} \ r\rightarrow \pm \infty.
\end{eqnarray}
All the results are regular. It is interesting that for $n=0$, the term that dominates at infinity is not the one that comes from Schwarzschild solution.

\begin{figure}[!htb]
	\includegraphics[scale=0.70]{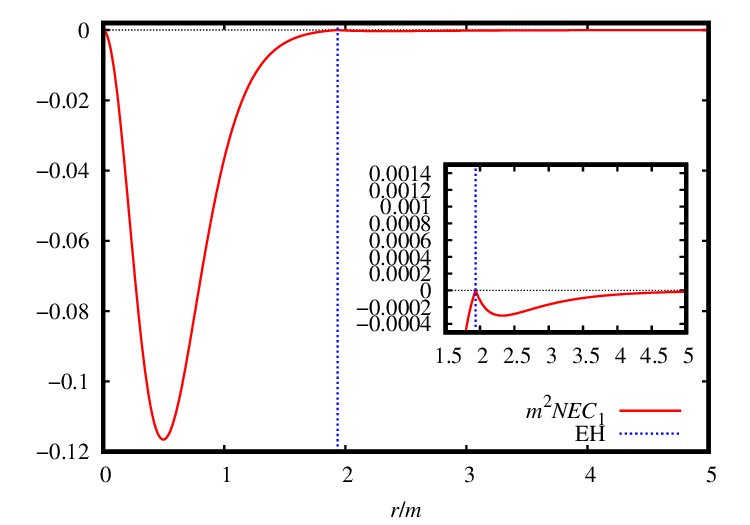}
	\includegraphics[scale=0.70]{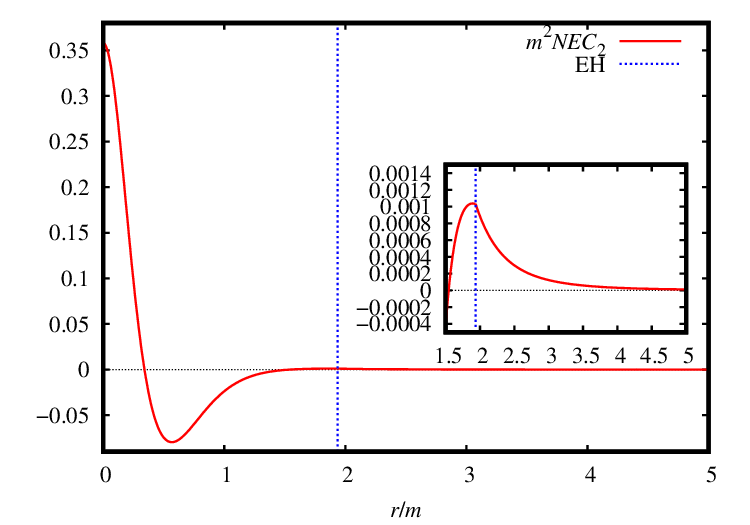}
	\caption{Null energy condition to the model \eqref{TRnewBBS} with $b=d=2a=m$ and $n=1$. As the functions are symmetric, we are considering only the positive part of the radial coordinate. The condition $NEC_1$ is always violated, while the condition $NEC_2$ is satisfied outside the event horizon.}\label{fig2TRnewBBS}
\end{figure}

The null energy condition is analyzed in Fig. \ref{fig2TRnewBBS}. As wee see, the functions $NEC_1$ and $NEC_2$ present negative values within the event horizon, besides that, $NEC_1$ is also violated outside, so the null energy condition is violated.
\subsubsection{Model $\Sigma(r)=\sqrt{\left(d^{2 n}+r^{2 n}\right)^{\frac{1}{n}}}$}
Let us consider the model
\begin{equation}
f(r)=1-\frac{2m}{\sqrt{r^2+a^2}}, \qquad \Sigma(r)=\sqrt{\left(d^{2 n}+r^{2 n}\right)^{\frac{1}{n}}}\,.\label{FourthRnewBBS}
\end{equation}
The area becomes
\begin{equation}
A=4 \pi\left(d^{2 n}+r^{2 n}\right)^{\frac{1}{n}}.
\end{equation}
This area is symmetric to $r\rightarrow -r$.
The area has only one real extreme value at $r\rightarrow 0$, which is a minimum. This metric has a similarity with the case before. In this case the area obeys
\begin{equation}
    \left.\frac{d^{2n-1}A}{dr^{2n-1}}\right|_{r=0}=0, \qquad \mbox{and} \qquad \left.\frac{d^{2n}A}{dr^{2n}}\right|_{r=0}>0.
\end{equation}
It means that the extreme point is a throat.

The Kretschmann scalar tends to a constant at $r\rightarrow0$, and behaves as $r^{-6}$ to $r\rightarrow\infty$. It means that the spacetime is curvature regular.

The condition $NEC_1$ for this spacetime is written as
\begin{eqnarray}
&&NEC_1\Longleftrightarrow
-\left(1-\frac{2m}{\sqrt{a^2+r^2}}\right)\frac{2 (2 n-1) d^{2 n} r^{2 n-2}}{\kappa ^2  \left(d^{2 n}+r^{2 n}\right)^2}
\geq 0,\quad \mbox{to} \quad f(r)>0,\\
&&NEC_1\Longleftrightarrow
\left(1-\frac{2m}{\sqrt{a^2+r^2}}\right)\frac{2 (2 n-1) d^{2 n} r^{2 n-2} }{\kappa ^2 \left(d^{2 n}+r^{2 n}\right)^2}
\geq 0, \quad \mbox{to} \quad f(r)<0.
\end{eqnarray}
Clearly $NEC_1$ is violated in all spacetime. With an appropriate choice of $n$, it is possible to satisfy $NEC_2$ outside the horizon.

\subsubsection{Model $\Sigma(r)=\frac{\sqrt{2 m^2+2 r^2}}{\sqrt{e^{\frac{2 m^2}{r^2}-\frac{m^4}{2
   r^4}}+1}}$}
\begin{figure}
    \centering
    \includegraphics[scale=0.7]{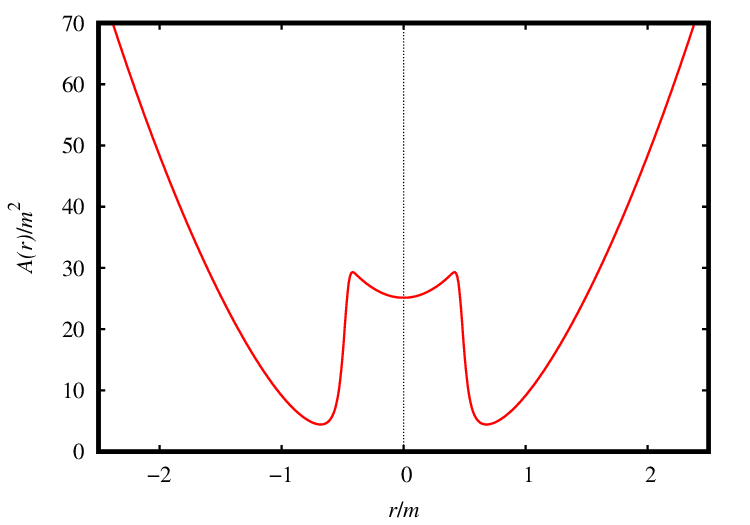}
    \caption{Behavior of the area as function of the radial coordinate.}
    \label{fig:AreaExtra2}
\end{figure}

Let us consider the model
\begin{equation}
f(r)=1-\frac{2m}{\sqrt{r^2+a^2}}, \qquad \Sigma(r)=\frac{\sqrt{2 m^2+2 r^2}}{\sqrt{e^{\frac{2 m^2}{r^2}-\frac{m^4}{2
   r^4}}+1}}\,.\label{LRnewBBS}
\end{equation}
The area becomes
\begin{equation}
A=\frac{8 \pi  \left(m^2+r^2\right)}{e^{\frac{2 m^2}{r^2}-\frac{m^4}{2
   r^4}}+1}.
\end{equation}
It is not possible to find the roots of $A'(r)$ analytically. In Fig. \ref{fig:AreaExtra2} we see the graphical representation of $A(r)$. There are five extremum points, two maximums and three minimums. It means that there are two anti-throats and three throats.
To $r\rightarrow 0$ the Kretschmann scalar tends to a constant, and to $r\rightarrow\infty$ behaves as $r^{-6}$. It means that there are no curvature singularity.

\begin{figure}
    \centering
    \includegraphics[scale=0.7]{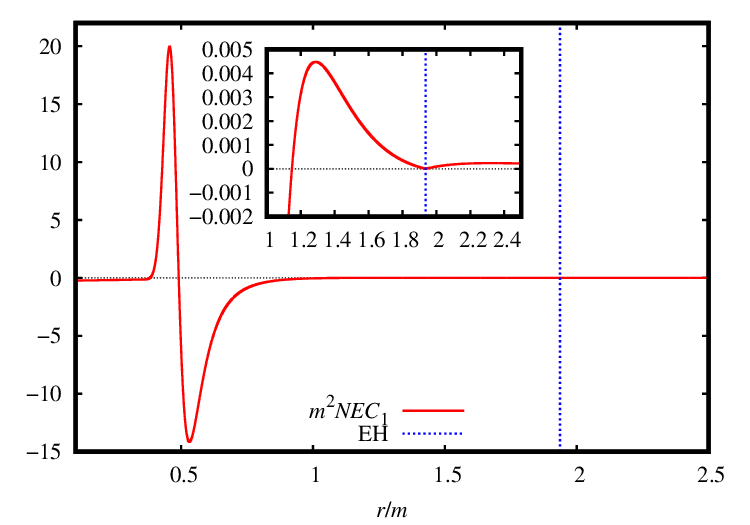}
    \includegraphics[scale=0.7]{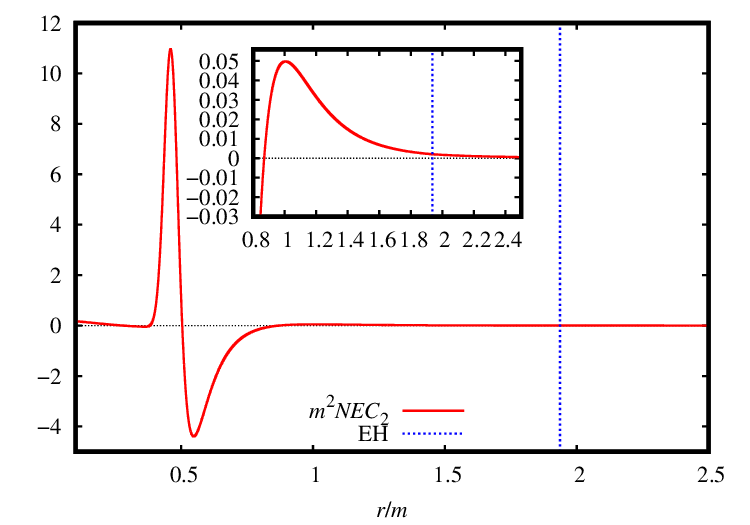}
   \caption{Null energy condition associated with \eqref{LRnewBBS} with $a=0.5m$. As the functions are symmetric to $r\rightarrow-r$, we consider only positive values of the radial coordinate.}
    \label{fig:ECExtra2}
\end{figure}

The null energy condition is shown in the Fig. \ref{fig:ECExtra2}. This condition is satisfied outside the event horizon, and violated inside the event horizon.
%==========================================
\section{Conclusion}\label{S:conclusion}
%==========================================
In this work, we build black-bounce spacetimes using Einstein equations. The proposed metrics present two regions with an infinity area and at least one throat between them. Some black-bounces have anti-throats, regions with a maximum in the area, and two throats. This type of behavior in the area appears in wormhole solutions \cite{Simpson:2018tsi, Boonserm:2018orb, Morris:1988cz}. The outcomes presented throughout this work can be succinctly summarized using Table \ref{t_results}.
\begin{table}[!htpb]
	\centering
	% definindo o tamanho da fonte para small
	% outros possíveis tamanhos: footnotesize, scriptsize		
		% redefinindo o espaçamento das colunas
		\setlength{\tabcolsep}{3pt} 
	\begin{tabular}{|c|c|c|c|c|c|c|c|c|c|c|c|c|c|c|c|c|c|c|}\hline
			\thead{Model ($\Sigma^2$)} & Regular & Throat & Anti-throat & $NEC_1$ & $NEC_2$  & Asymptotically Minkowski ($r\rightarrow +\infty$)\\ \hline
			\thead{$r^2e^{b^2/r^2}$ } & No & One & No & Nowhere & Outside & Yes \\ \hline
			\thead{$r^2 e^{\frac{b^2}{m r+c_3}}$} & No & Two & One & Nowhere & Nowhere & Yes \\ \hline
			\thead{ $\left(b^2+r^2\right) e^{\frac{b^2}{m r+c_3}}$} & No & One & No & Nowhere & Nowhere & Yes \\ \hline
			\thead{ $\frac{m^2}{e^{-\frac{m^2\left(m-8r\right)}{8r^3}}-1}$} & No & One & No & Outside & Outside & Yes \\ \hline
			\thead{$\left(b^2+r^2\right) e^{\frac{b^2}{c_3+r^2}}$} & Yes & Three & Two & Nowhere & Nowhere & Yes \\ \hline
			\thead{$b^2+r^2 \cos \left(\frac{d}{\sqrt{d^2+r^2}}\right)$} & Yes & One & No & Nowhere & Outside & Yes \\ \hline
			\thead{$b^2+\frac{r^{2n+2}}{d^{2 n}+r^{2 n}}$} & Yes & One & No & Nowhere & Outside & Yes \\ \hline      \thead{$\left(d^{2 n}+r^{2 n}\right)^{\frac{1}{n}}$} & Yes & One & No & Nowhere & Outside & Yes   \\ \hline
            \thead{$\frac{2 m^2+2 r^2}{e^{\frac{2 m^2}{r^2}-\frac{m^4}{2
   r^4}}+1}$} & Yes & Three & Two & Outside & Outside & Yes \\ \hline
		\end{tabular} 
	\caption{Summary of properties for each model. Regarding the number of throats or anti-throats, we are specifying the maximum number each model can have, as this number may decrease depending on the parameters. About the energy conditions, we are specifying the region in spacetime where they are satisfied, which can be within the event horizon, outside the event horizon, throughout the entire spacetime, or in no region at all.}
	\label{t_results}
\end{table} 

Based on the curvature regularity of the spacetime, there are two types of spacetimes; regular and singular. In both cases, the throat is located inside the event horizon. The regions with infinite areas are located within the event horizon in the singular spacetimes. This infinity represents a singularity in the curvature invariants. At least one of the infinities in the area represents a curvature singularity. We also have only one throat and, in the case \eqref{snewBBS}, there is also an anti-throat. In regular cases, both regions of infinite area occur at $r\rightarrow \pm \infty$. Unlike the other metrics, the regions with infinite areas do not present singularities in the curvature invariants. Depending on the parameters, the regular black-bounces have more than one throat; between these throats, there is an anti-throat. Wormhole solutions with anti-throats can appear when you have a non-minimal coupling with the scalar field \cite{Ibadov:2020btp}.

All the energy conditions are related to the null energy conditions. So, if the null energy condition is violated, all other energy conditions also are violated. Almost all spacetimes that were built in this work present the behavior $\Sigma''(r)/\Sigma(r)>0$, which guarantees that at least one of the inequalities of the null condition is violated. However, the cases \eqref{LSnewBBS} and \eqref{LRnewBBS} satisfy the null energy condition outside the event horizon. In these two cases, only the dominant energy condition is continually violated. In contrast, the remaining energy conditions are satisfied outside the event horizon, the only region causally connected to observers.

The behavior of these spacetimes is similar to existing solutions where alternative theories of gravitation or scalar fields are considered. In this way, we believe it is possible to find the content of matter that generates them.

Usually, black-bounce solutions can be obtained by coupling gravitational theory with nonlinear electrodynamics plus a phantom scalar field \cite{Bronnikov:2021uta,Canate:2022gpy}. The phantom field is responsible for violating the null energy condition. There are also solutions that are partially phantom. Therefore, it is possible that the spacetimes presented in this work are solutions for cases in which we have electrodynamics with a phantom scalar field, for cases in which the null energy condition is everywhere violated, or partially phantom, for cases in which the null energy condition is violated only in certain regions of spacetime.

In future works, we hope to find the source of these metrics. We also expect be able to study the thermodynamics. The problem with thermodynamics is related to the material content that gives rise to these metrics. We also hope to study the behavior of null geodesics since these geodesics are associated with the black hole shadows.

%==========================================

%==========================================
\section*{Acknowledgements}
The authors would like to thank Matt Visser for useful comments about the work. MER  thanks Conselho Nacional de Desenvolvimento Cient\'ifico e Tecnol\'ogico - CNPq, Brazil  for partial financial support. This study was financed in part by the Coordena\c{c}\~ao de Aperfei\c{c}oamento de Pessoal de N\'ivel Superior - Brasil (CAPES) - Finance Code 001.

%==========================================
\appendix
\section{Carter-Penrose diagrams}\label{Appen_A}
\begin{figure}
    \centering
    \subfigure[]{\includegraphics[scale=0.2]{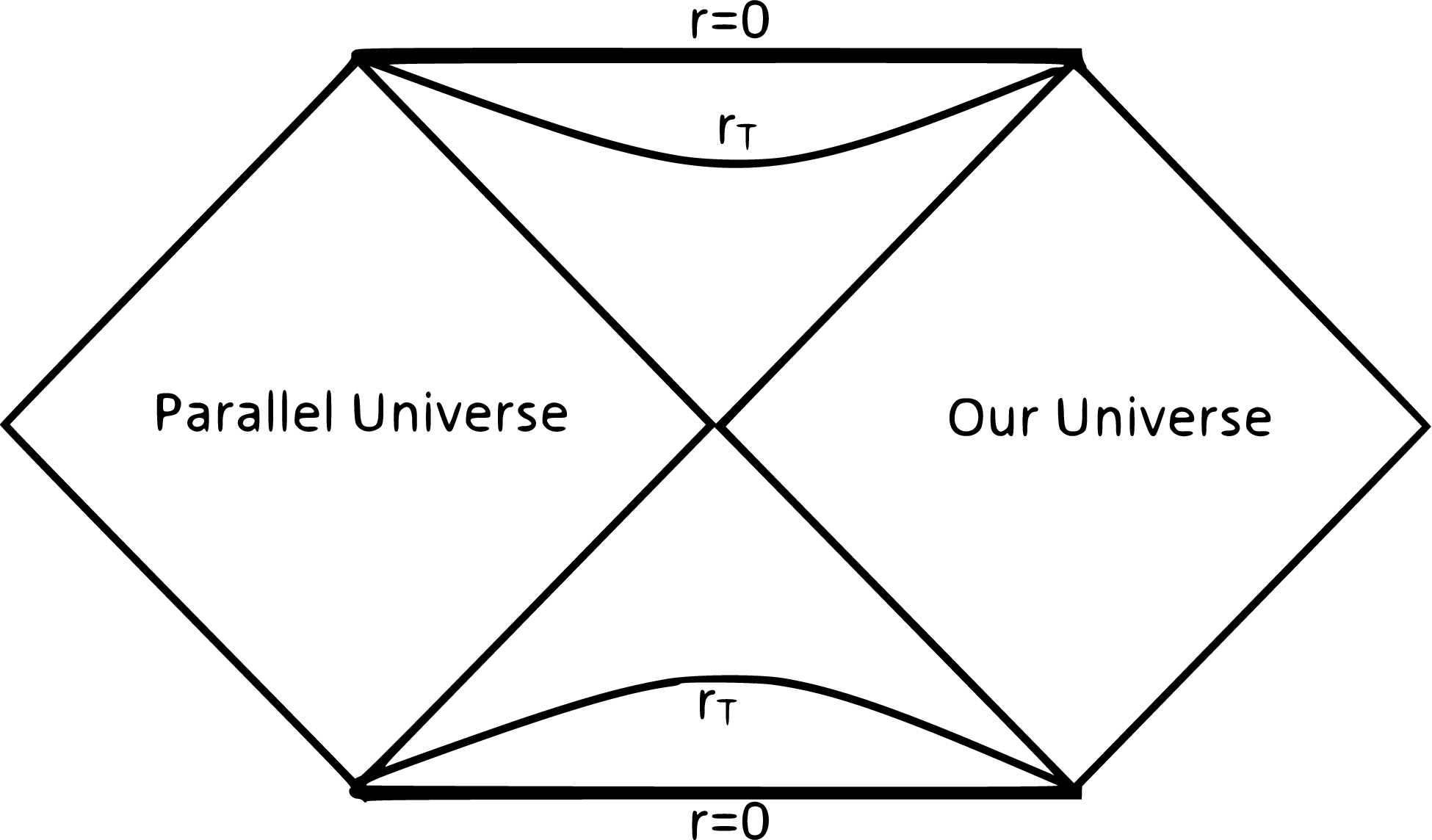}}
    \subfigure[]{\includegraphics[scale=0.2]{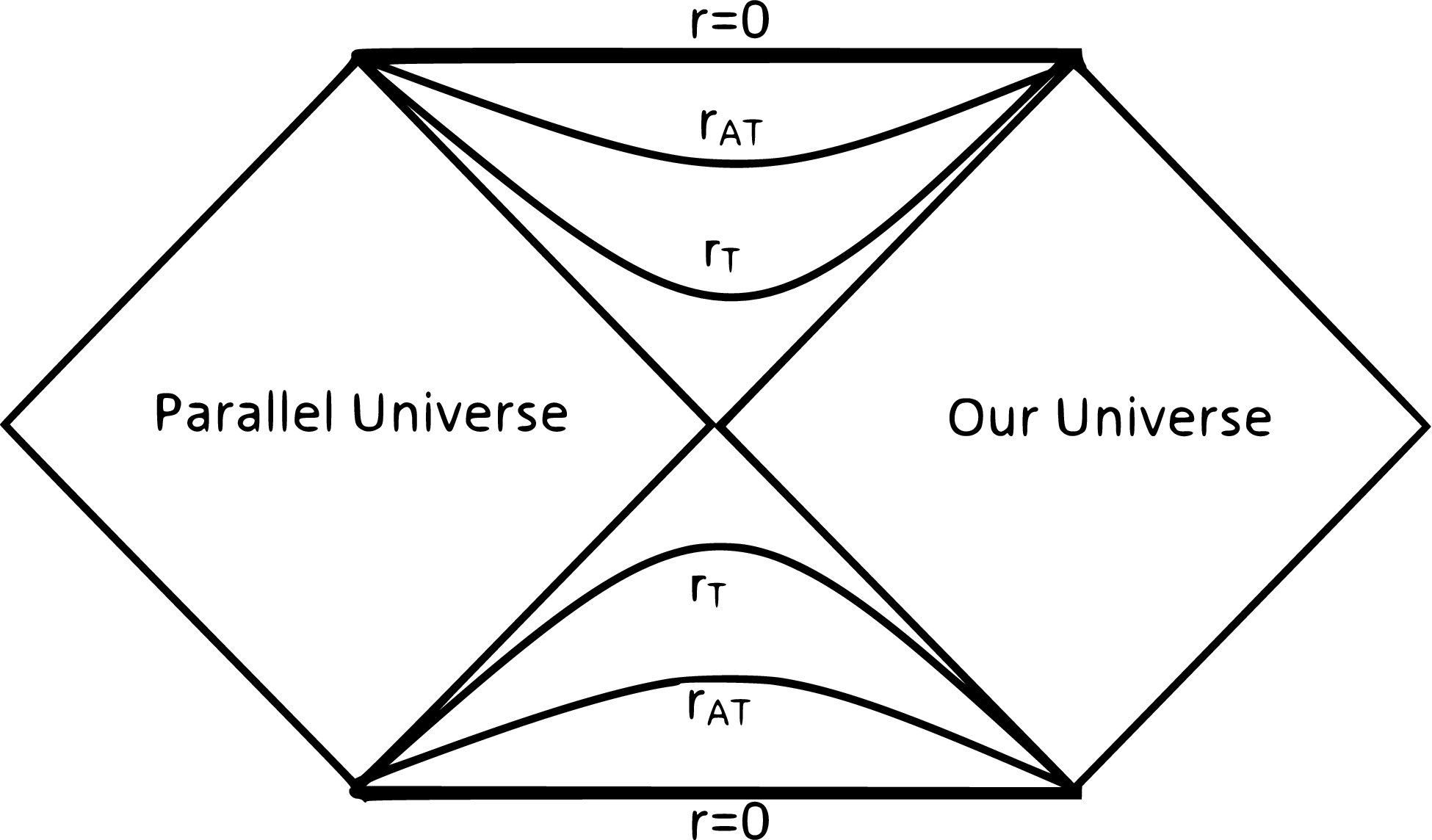}}
    \subfigure[]{\includegraphics[scale=0.15]{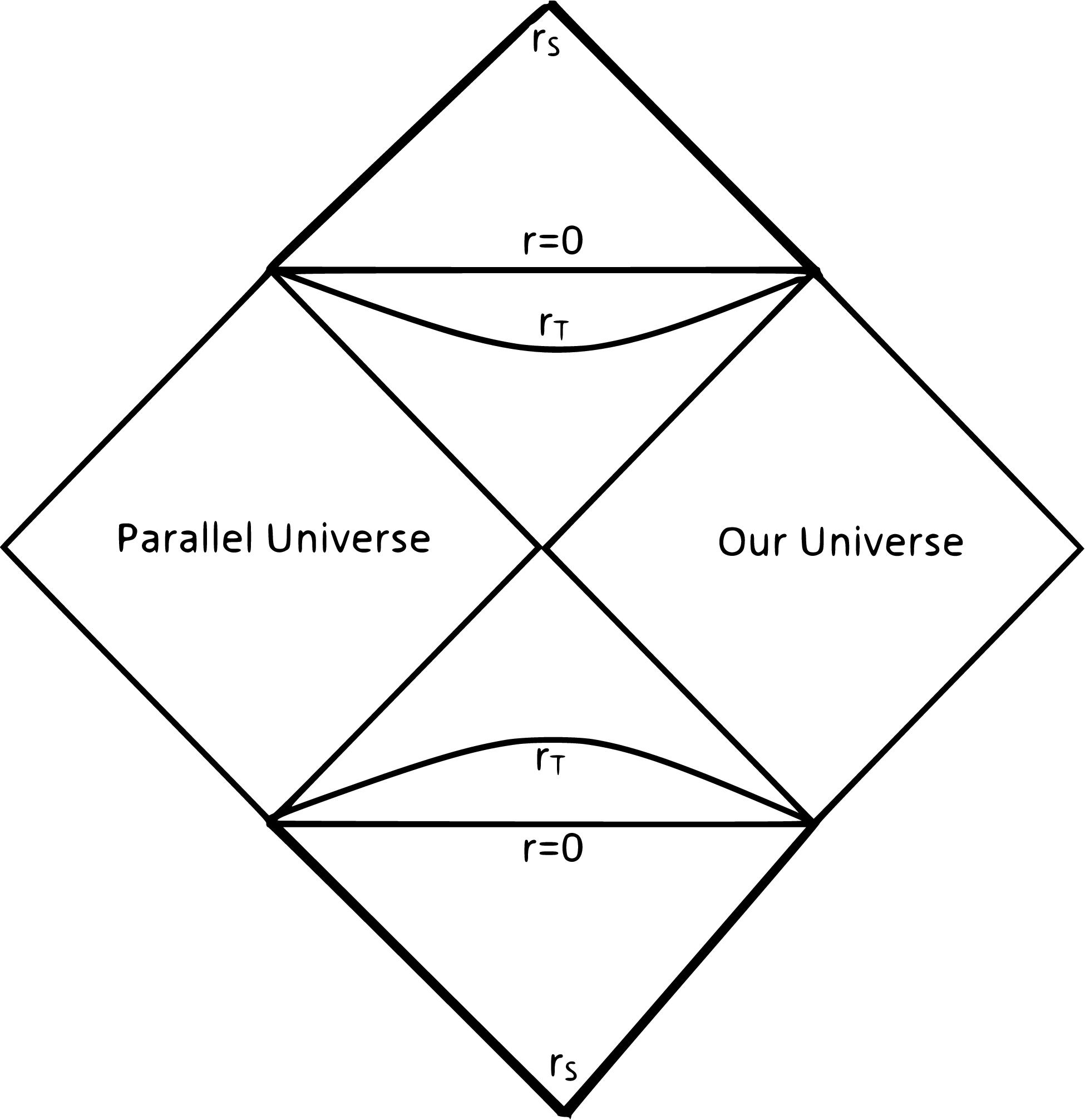}}\hspace{1cm}
    \subfigure[]{\includegraphics[scale=0.2]{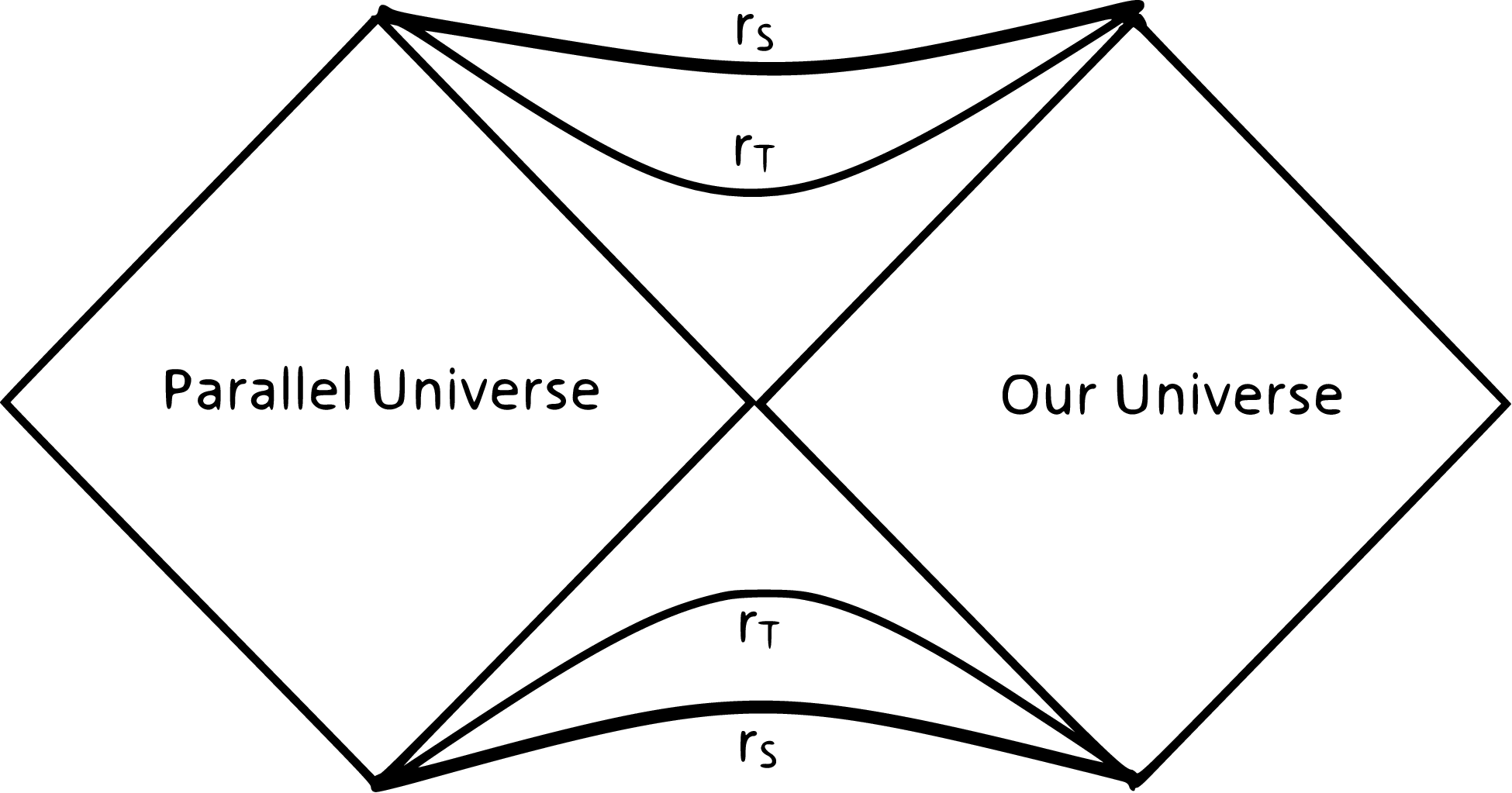}}
    \caption{Carter-Penrose diagrams for the black-bounces with singularities. In (a), we have the diagram for the case where the throat is located inside the event horizon, and there is a singularity at $r=0$. In (b), we have the case where there is a throat and an anti-throat located within the event horizon, and a throat at the same location as the singularity, at $r=0$. In (c), the Carter-Penrose diagram describes the case where we have a throat within the event horizon, and a singularity beyond the point $r=0$. In (d), we have the Carter-Penrose diagram to the case where there is a throat within the event horizon, and a singularity before $r=0$.}
    \label{fig:Penrose_Singular}
\end{figure}

In this appendix, we present the Carter-Penrose diagrams describing the causal structure for certain cases of black-bounces.

The black-bounce described by \eqref{fnewBBS} features a singularity at $r=0$ and a throat at $r=b$. These structures are located within the event horizon, and as a result, the causal structure is depicted by the Carter-Penrose diagram in Fig. \ref{fig:Penrose_Singular} (a). In the model \eqref{snewBBS}, within the horizon, there exists a throat and an anti-throat. Furthermore, there is a second throat located at the same point as the singularity, $r=0$, rendering it singular. Consequently, only one of the throats is traversable. This causal structure is depicted in Fig. \ref{fig:Penrose_Singular} (b). In the model \eqref{tnewBBS}, there is only one throat that, depending on the chosen parameters, can be located inside or outside the horizon. The point $r=0$ doesn't present any singularity, it's just an arbitrary point in the manifold, but there is a singularity at $r=-c_3/m$. The causal structure of this model, when considering the throat is located inside the horizon, is represented in Fig. \ref{fig:Penrose_Singular} (c). The model \eqref{LSnewBBS} features only one throat located at $r=3m/16$ and a singularity at $r=m/8$. Thus, the singularity is situated before $r=0$, but it still lies within the event horizon. The Carter-Penrose diagram that illustrates this causal structure is given by Fig. \ref{fig:Penrose_Singular} (d).

\begin{figure}
    \centering
    \subfigure[]{\includegraphics[scale=0.235]{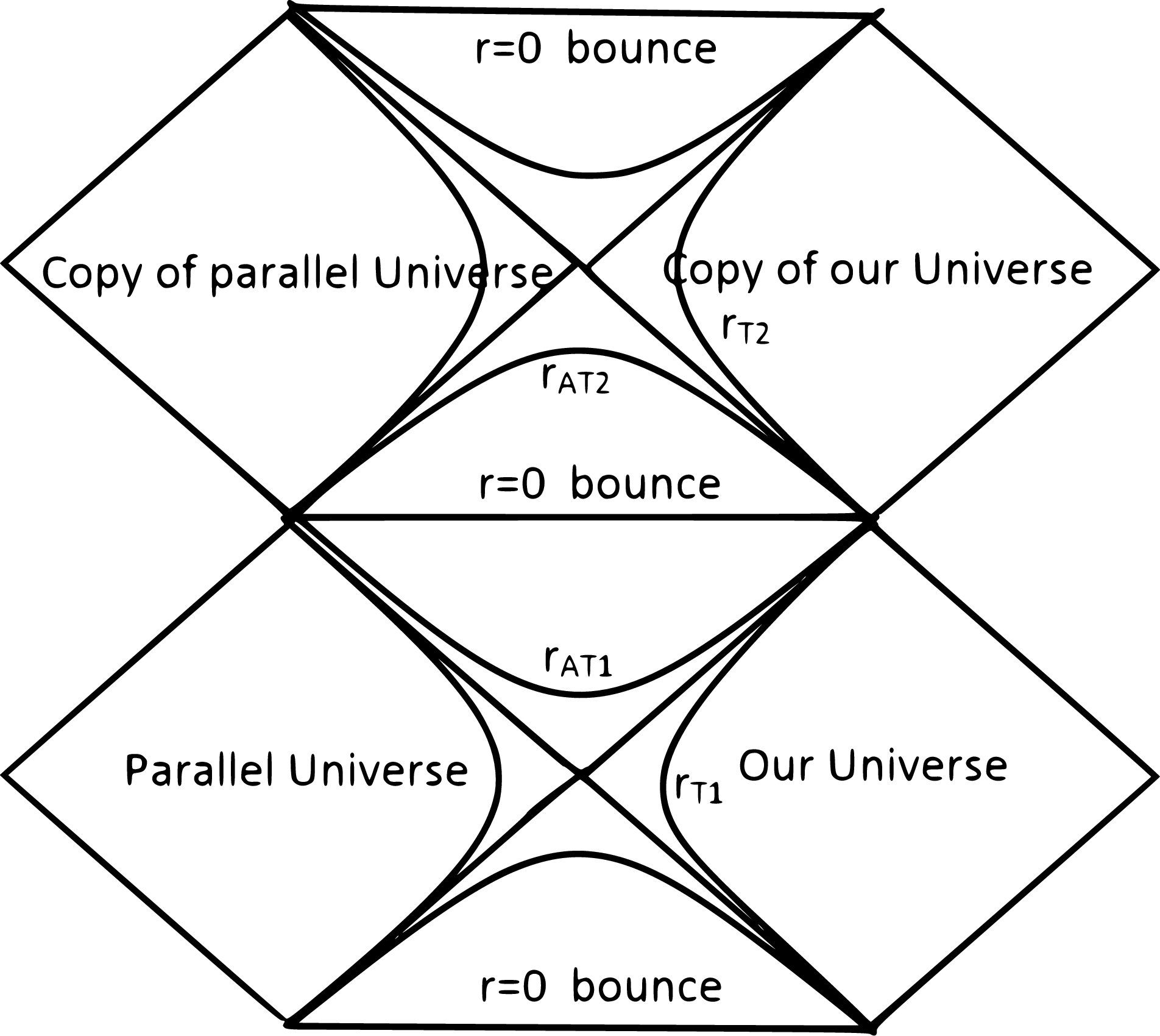}}
    \subfigure[]{\includegraphics[scale=0.2]{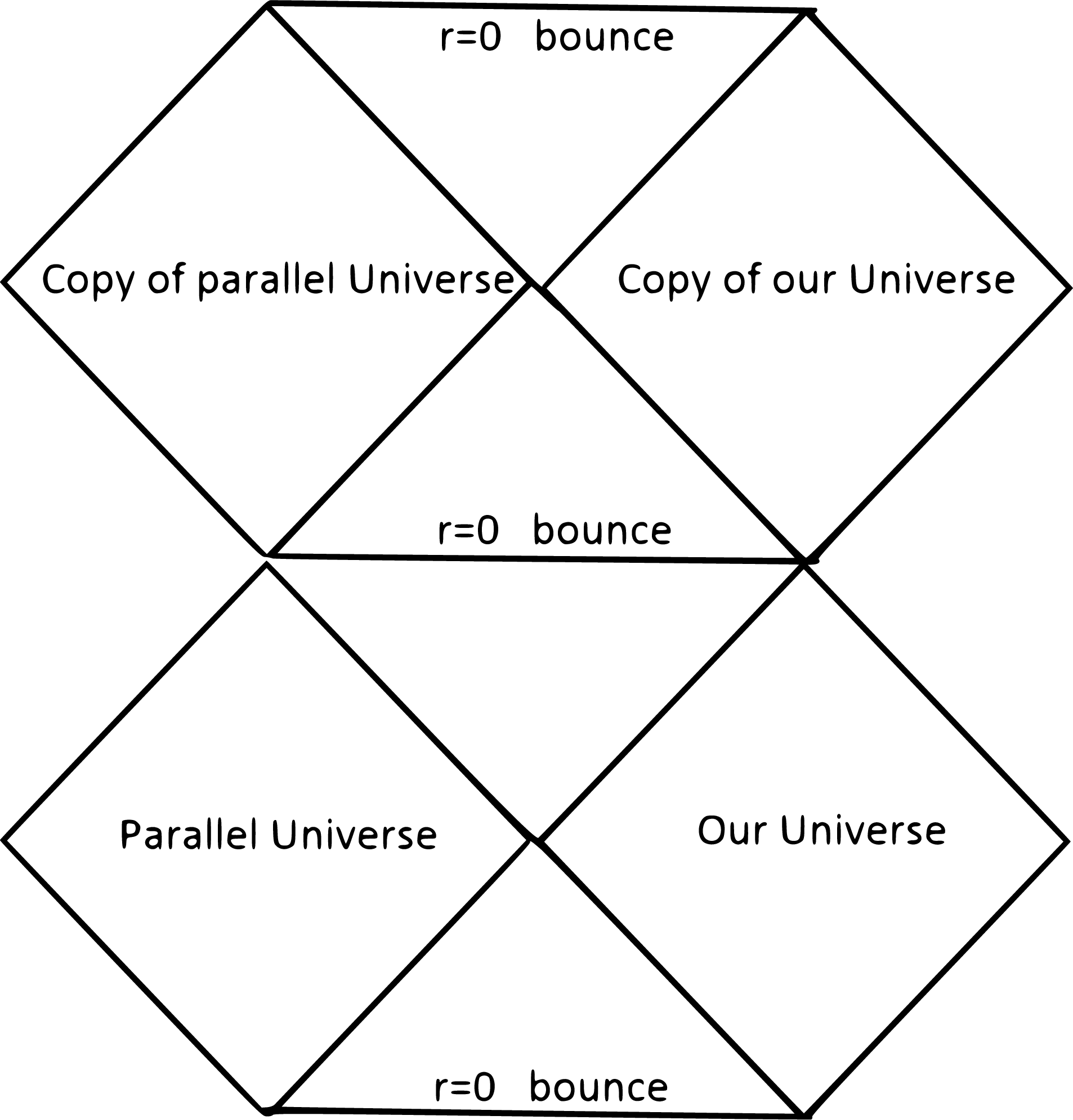}}
    \subfigure[]{\includegraphics[scale=0.2]{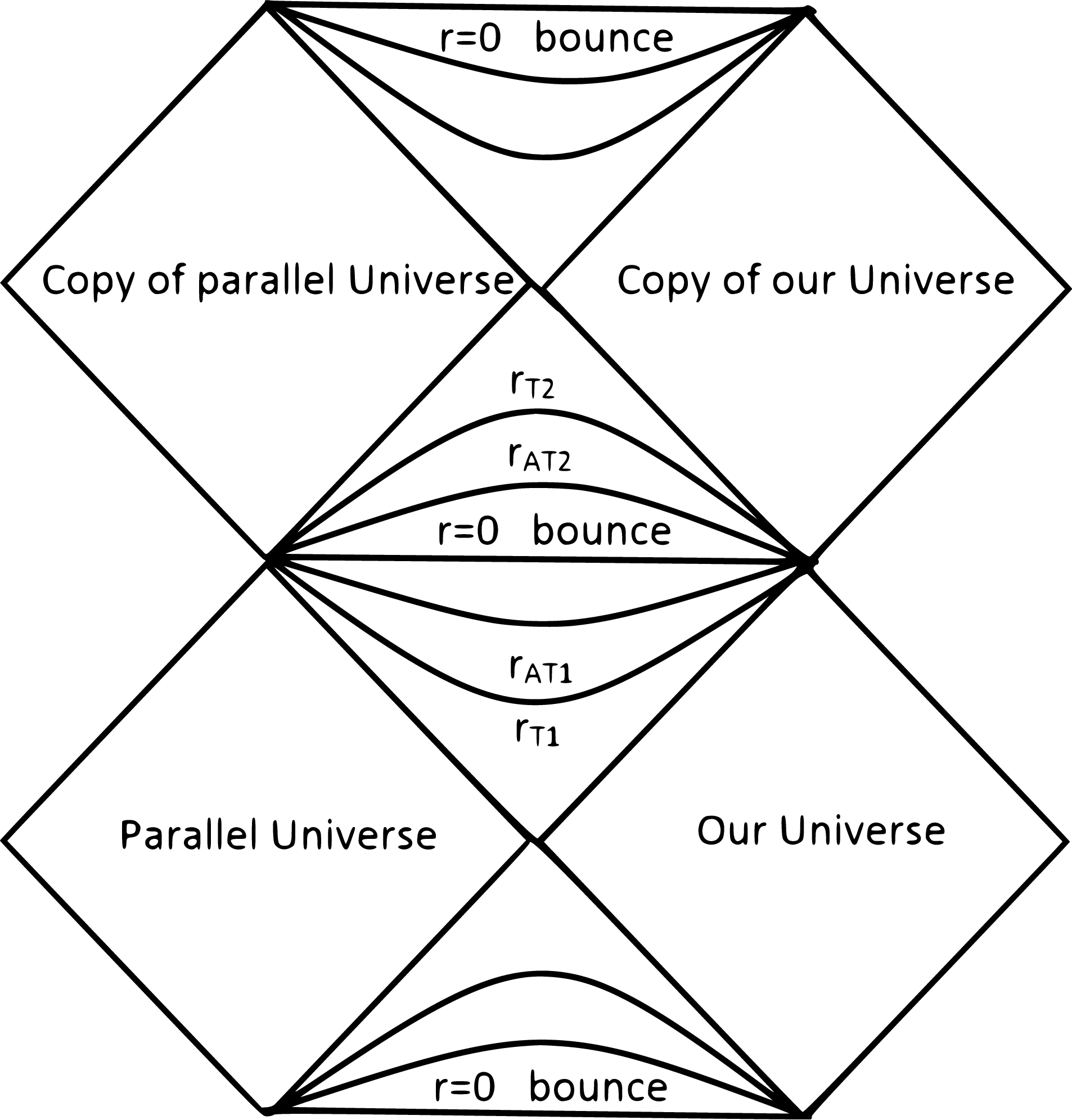}}
    \caption{Carter-Penrose diagrams illustrate the regular black-bounce in various scenarios. In (a), we describe a configuration where a throat exists outside the event horizon, an anti-throat within the event horizon, and a bounce occurs at $r=0$. This behavior repeats symmetrically in the negative part of the radial coordinate. In (b), we consider a case where there is only a bounce at $r=0$. In (c), the Carter-Penrose diagram represents a scenario with both a throat and an anti-throat inside the event horizon and a bounce at $r=0$. This behavior mirrors itself in the negative part of the radial coordinate.}
    \label{fig:Penrose_Regular}
\end{figure}

Depending on the parameter choices, the regular model \eqref{fRnewBBS} can exhibit up to three throats and two anti-throats. In the case where the number of throats is at its maximum, one of them is located at $r=0$, thus forming a "bounce", while the two anti-throats are located within the horizon, and the remaining two throats are situated outside the horizon. In this way, the Carter-Penrose diagram that best describes the causal structure of this spacetime is the one shown in Fig. \ref{fig:Penrose_Regular} (a). The regular models \eqref{SRnewBBS}, \eqref{TRnewBBS}, and \eqref{FourthRnewBBS} basically present the same causal structure, since all three spacetimes have only one throat, located at $r=0$, which is positioned between the event horizons, Fig. \ref{fig:Penrose_Regular} (b). Similar to the model \eqref{fRnewBBS}, the model \eqref{LRnewBBS} can also have up to three throats and two anti-throats. However, in this case, all the throats are located within the event horizon, and thus, the causal structure is represented by the Carter-Penrose diagram in Fig. \ref{fig:Penrose_Regular} (c).
%==========================================

%==========================================

\begin{thebibliography}{99}
%\cite{Stoica:2014tpa}
\bibitem{Stoica:2014tpa}
O.~C.~Stoica,
``The Geometry of Black Hole singularities'',
Adv. High Energy Phys. \textbf{2014}, 907518 (2014),
%doi:10.1155/2014/907518
[arXiv:1401.6283 [gr-qc]].
%3 citations counted in INSPIRE as of 27 Mar 2022


%\cite{Ong:2020xwv}
\bibitem{Ong:2020xwv}
Y.~C.~Ong,
``Space\textendash{}time singularities and cosmic censorship conjecture: A Review with some thoughts'',
Int. J. Mod. Phys. A \textbf{35}, no.14, 14 (2020),
%doi:10.1142/S0217751X20300070
[arXiv:2005.07032 [gr-qc]].
%8 citations counted in INSPIRE as of 27 Mar 2022

%==========================================

%\cite{Bronnikov:2012wsj}
\bibitem{Bronnikov:2012wsj}
K.~A.~Bronnikov and S.~G.~Rubin,
``Black Holes, Cosmology and Extra Dimensions'', World
Scientific, Singapore (2013).
%2 citations counted in INSPIRE as of 10 Aug 2020

\bibitem{wal} R. M. Wald, ``General Relativity'', The University of Chicago Press, Chicago (1984).



%\cite{Tipler:1978zz}
\bibitem{Tipler:1978zz}
F.~J.~Tipler,
``Energy conditions and spacetime singularities'',
Phys. Rev. D \textbf{17}, 2521-2528 (1978).
%doi:10.1103/PhysRevD.17.2521
%142 citations counted in INSPIRE as of 27 Mar 2022


%\cite{Lobo:2020ffi}
\bibitem{Lobo:2020ffi}
F.~S.~N.~Lobo, M.~E.~Rodrigues, M.~V.~d.~S.~Silva, A.~Simpson and M.~Visser,
``Novel black-bounce spacetimes: wormholes, regularity, energy conditions, and causal structure'',
Phys. Rev. D \textbf{103}, no.8, 084052 (2021),
%doi:10.1103/PhysRevD.103.084052
[arXiv:2009.12057 [gr-qc]].
%35 citations counted in INSPIRE as of 26 Mar 2022


%====================================================

%\cite{Rendall:2005nf}
\bibitem{Rendall:2005nf}
A.~D.~Rendall,
%``The Nature of spacetime singularities,''
100 Years Of Relativity : space-time structure: Einstein and beyond, 76-92 (2005),
%doi:10.1142/9789812700988\_0003
[arXiv:gr-qc/0503112 [gr-qc]].
%20 citations counted in INSPIRE as of 27 Mar 2022


\bibitem{din} R. D'Inverno, ``Introducing Einstein's Relativity'', Oxford University Press, New York (1998).



%\cite{Ansoldi:2008jw}
\bibitem{Ansoldi:2008jw}
S.~Ansoldi,
``Spherical black holes with regular center: A Review of existing models including a recent realization with Gaussian sources'',
[arXiv:0802.0330 [gr-qc]].
%191 citations counted in INSPIRE as of 27 Mar 2022

\bibitem{Bardeen} 
J. M. Bardeen,{\it{ Non-singular general relativistic gravitational collapse}}, in Proceedings of the International Conference GR5, Tbilisi, U.S.S.R. (1968).

\bibitem{Beato1}
E.~Ayon-Beato and A.~Garcia,
``The Bardeen model as a nonlinear magnetic monopole'',
Phys. Lett. B \textbf{493}, 149-152 (2000),
%doi:10.1016/S0370-2693(00)01125-4
[arXiv:gr-qc/0009077 [gr-qc]].
%390 citations counted in INSPIRE as of 27 Mar 2022


%====================================================
\bibitem{Fan-Wang}
Z.~Y.~Fan and X.~Wang, ``Construction of Regular Black Holes in General Relativity'',
Phys. Rev. D {\bf 94}, no.12, 124027 (2016),
[arXiv:1610.02636 [gr-qc]].

\bibitem{Zaslavskii}
O.~B.~Zaslavskii, ``Regular black holes and energy conditions'', Phys. Lett. B {\bf 688}, 278-280 (2010),
[arXiv:1004.2362 [gr-qc]].

\bibitem{Rodrigues:2015}
M.~E.~Rodrigues, E.~L.~B.~Junior, G.~T.~Marques and V.~T.~Zanchin,
``Regular black holes in $f(R)$ gravity coupled to nonlinear electrodynamics'',
Phys. Rev. D \textbf{94} (2016) no.2, 024062,
%doi:10.1103/PhysRevD.94.024062
[arXiv:1511.00569 [gr-qc]].
%30 citations counted in INSPIRE as of 15 Aug 2020

\bibitem{Rodrigues:2016}
M.~E.~Rodrigues, J.~C.~Fabris, E.~L.~B.~Junior and G.~T.~Marques,
``Generalisation for regular black holes on general relativity to $f(R)$ gravity'',
Eur. Phys. J. C \textbf{76} (2016) no.5, 250,
%doi:10.1140/epjc/s10052-016-4085-x
[arXiv:1601.00471 [gr-qc]].
%22 citations counted in INSPIRE as of 15 Aug 2020

\bibitem{Rodrigues:2017}
M.~E.~Rodrigues, E.~L.~B.~Junior and M.~V.~de S.~Silva,
``Using dominant and weak energy conditions for building new classes of regular black holes'',
JCAP \textbf{02} (2018), 059,
%doi:10.1088/1475-7516/2018/02/059
[arXiv:1705.05744 [physics.gen-ph]].
%9 citations counted in INSPIRE as of 15 Aug 2020

\bibitem{Rodrigues:2018}
M.~E.~Rodrigues and M.~V.~d.~Silva,
``Bardeen Regular Black Hole With an Electric Source'',
JCAP \textbf{06} (2018), 025,
%doi:10.1088/1475-7516/2018/06/025
[arXiv:1802.05095 [gr-qc]].
%14 citations counted in INSPIRE as of 15 Aug 2020

\bibitem{Rodrigues:2019}
M.~E.~Rodrigues and M.~V.~de S.~Silva,
``Regular multi-horizon black holes in $f(G)$ gravity with nonlinear electrodynamics'',
Phys. Rev. D \textbf{99} (2019) no.12, 124010,
%doi:10.1103/PhysRevD.99.124010
[arXiv:1906.06168 [gr-qc]].
%3 citations counted in INSPIRE as of 15 Aug 2020

\bibitem{Silva:2018}
M.~V.~d.~Silva and M.~E.~Rodrigues,
``Regular black holes in $f(G)$ gravity'',
Eur. Phys. J. C \textbf{78} (2018) no.8, 638,
%doi:10.1140/epjc/s10052-018-6122-4
[arXiv:1808.05861 [gr-qc]].
%8 citations counted in INSPIRE as of 15 Aug 2020

\bibitem{Junior:2020}
E.~L.~B.~Junior, M.~E.~Rodrigues and M.~V.~de Sousa Silva,
``Regular black holes in Rainbow Gravity'',
Nucl. Phys. B \textbf{961}, 115244 (2020),
%doi:10.1016/j.nuclphysb.2020.115244
[arXiv:2002.04410 [gr-qc]].
%14 citations counted in INSPIRE as of 11 Oct 2023



%\cite{Bronnikov:2017tnz}
\bibitem{Bronnikov:2017tnz}
K.~A.~Bronnikov,
``Comment on \textquotedblleft{}Construction of regular black holes in general relativity\textquotedblright{}'',
Phys. Rev. D \textbf{96}, no.12, 128501 (2017)
%doi:10.1103/PhysRevD.96.128501
[arXiv:1712.04342 [gr-qc]].
%28 citations counted in INSPIRE as of 24 Nov 2021




%\cite{Capozziello:2014bqa}
\bibitem{Capozziello:2014bqa}
S.~Capozziello, F.~S.~N.~Lobo and J.~P.~Mimoso,
``Generalized energy conditions in Extended Theories of Gravity'',
Phys. Rev. D \textbf{91} (2015) no.12, 124019
%doi:10.1103/PhysRevD.91.124019
[arXiv:1407.7293 [gr-qc]].
%61 citations counted in INSPIRE as of 25 Aug 2020

\bibitem{NED2}
E. Ay\'on-Beato and A. Garc\'{\i}a, ``New regular black hole solution 
from nonlinear electrodynamics'', Phys.Lett. B {\bf 464}, 25 (1999) 
[arXiv:9911174 [hep-th]].

\bibitem{NED3}
E. Ay\'on-Beato and A. Garc\'\i a, ``Regular black hole in general 
relativity coupled to nonlinear electrodynamics'', Phys. Rev. Lett. {\bf 80}, 
5056 (1998) [arXiv:9911046 [gr-qc]].

\bibitem{NED4}
K. A. Bronnikov, ``Regular magnetic black holes 
and monopoles from nonlinear electrodynamics'', Phys. Rev. D {\bf 63}, 044005 
(2001) [arXiv:0006014 [gr-qc]].

\bibitem{NED5}
I. Dymnikova, ``Regular electrically charged structures in nonlinear 
electrodynamics coupled to general relativity'', Classical Quantum Gravity 
{\bf 21}, 4417 (2004), [arXiv:0407072 [gr-qc]].


%\cite{Bronnikov:2006fu}
\bibitem{Bronnikov:2006fu}
K.~A.~Bronnikov, V.~N.~Melnikov and H.~Dehnen,
``Regular black holes and black universes'',
Gen. Rel. Grav. \textbf{39}, 973-987 (2007)
%doi:10.1007/s10714-007-0430-6
[arXiv:gr-qc/0611022 [gr-qc]].
%109 citations counted in INSPIRE as of 24 Nov 2021


%\cite{Hollenstein:2008hp}
\bibitem{Hollenstein:2008hp}
L.~Hollenstein and F.~S.~N.~Lobo,
``Exact solutions of $f(R)$ gravity coupled to nonlinear electrodynamics'',
Phys. Rev. D \textbf{78} (2008), 124007,
%doi:10.1103/PhysRevD.78.124007
[arXiv:0807.2325 [gr-qc]].
%67 citations counted in INSPIRE as of 25 Aug 2020

\bibitem{NED10}
L. Balart and E. C. Vagenas, ``Regular black holes with a nonlinear 
electrodynamics source'', Phys. Rev. D {\bf 90}, 124045 (2014),
[arXiv:1408.0306 [gr-qc]].


\bibitem{bambi} 
C. Bambi, L. Modesto, ``Rotating regular black holes'', 
Phys. Lett. B \textbf{721} (2013), 329-334,
[arXiv:1302.6075 [gr-qc]].

\bibitem{neves} 
J.~C.~S.~Neves, A. Saa, ``Regular rotating black holes and the weak energy condition'', 
Phys. Lett. B \textbf{734} (2014), 44-48,
[arXiv:1402.2694 [gr-qc]].

\bibitem{toshmatov} 
B. Toshmatov, B. Ahmedov, A. Abdujabbarov, Z. Stuchlik, ``Rotating Regular Black Hole Solution'', 
Phys. Rev. D \textbf{89} (2014) no. 10, 104017, 
[arXiv:1404.6443 [gr-qc]].


\bibitem{DYM} 
I. Dymnikova, E. Galaktionov, ``Regular rotating electrically charged black holes and solitons in non-linear electrodynamics minimally coupled to gravity'', 
Class. Quant. Grav. \textbf{32} (2015) no. 16, 165015,
[arXiv:1510.01353 [gr-qc]].

\bibitem{ramon} 
R. Torres, F. Fayos, ``On regular rotating black holes'', 
Gen. Rel. Grav. \textbf{49} (2017) no. 1, 2, Quant. Grav. \textbf{32} (2015) no. 16, 165015,
[arXiv:1611.03654 [gr-qc]].

\bibitem{berej} 
W. Berej, J. Matyjasek, D. Tryniecki, M. Woronowicz, ``Regular black holes in quadratic gravity'', 
Gen. Rel. Grav. \textbf{38} (2006) 885-906,
[arXiv:0606185 [hep-th]].




%=====================================================
%\cite{Simpson:2018tsi}
\bibitem{Simpson:2018tsi}
A.~Simpson and M.~Visser,
``Black-bounce to traversable wormhole'',
JCAP \textbf{02} (2019), 042
%doi:10.1088/1475-7516/2019/02/042
[arXiv:1812.07114 [gr-qc]].
%13 citations counted in INSPIRE as of 30 Jul 2020

%\cite{Simpson:2021vxo}
\bibitem{Simpson:2021vxo}
A.~Simpson,
``From black-bounce to traversable wormhole, and beyond'',
[arXiv:2110.05657 [gr-qc]].
%2 citations counted in INSPIRE as of 29 Mar 2022

%\cite{Bronnikov:2021uta}
\bibitem{Bronnikov:2021uta}
K.~A.~Bronnikov and R.~K.~Walia,
``Field sources for Simpson-Visser spacetimes'',
Phys. Rev. D \textbf{105}, no.4, 044039 (2022),
%doi:10.1103/PhysRevD.105.044039
[arXiv:2112.13198 [gr-qc]].
%4 citations counted in INSPIRE as of 29 Mar 2022


%\cite{Canate:2022gpy}
\bibitem{Canate:2022gpy}
P.~Ca\~nate,
``Black bounces as magnetically charged phantom regular black holes in Einstein-nonlinear electrodynamics gravity coupled to a self-interacting scalar field'',
Phys. Rev. D \textbf{106}, no.2, 024031 (2022),
%doi:10.1103/PhysRevD.106.024031
[arXiv:2202.02303 [gr-qc]].
%13 citations counted in INSPIRE as of 11 Oct 2023




\bibitem{Simpson:2019cer}
A.~Simpson, P.~Mart\'in-Moruno and M.~Visser,
``Vaidya spacetimes, black-bounces, and traversable wormholes'',\\
Class. Quant. Grav. \textbf{36} (2019) no.14, 145007,
%doi:10.1088/1361-6382/ab28a5
[arXiv:1902.04232 [gr-qc]].
%10 citations counted in INSPIRE as of 14 Aug 2020

\bibitem{Lobo:2020kxn}
F.~S.~N.~Lobo, A.~Simpson and M.~Visser,
``Dynamic thin-shell black-bounce traversable wormholes'',\\
Phys. Rev. D \textbf{101} (2020) no.12, 124035,
%doi:10.1103/PhysRevD.101.124035
[arXiv:2003.09419 [gr-qc]].
%5 citations counted in INSPIRE as of 14 Aug 2020








%\cite{Tsukamoto:2021caq}
\bibitem{Tsukamoto:2021caq}
N.~Tsukamoto,
``Gravitational lensing by two photon spheres in a black-bounce spacetime in strong deflection limits'',
Phys. Rev. D \textbf{104}, no.6, 064022 (2021),
%doi:10.1103/PhysRevD.104.064022
[arXiv:2105.14336 [gr-qc]].
%17 citations counted in INSPIRE as of 26 Mar 2022

%\cite{Olmo:2021piq}
\bibitem{Olmo:2021piq}
G.~J.~Olmo, D.~Rubiera-Garcia and D.~S.~C.~G\'omez,
``New light rings from multiple critical curves as observational signatures of black hole mimickers'',
Phys. Lett. B \textbf{829}, 137045 (2022),
%doi:10.1016/j.physletb.2022.137045
[arXiv:2110.10002 [gr-qc]].
%13 citations counted in INSPIRE as of 11 Oct 2023

%\cite{Guerrero:2022qkh}
\bibitem{Guerrero:2022qkh}
M.~Guerrero, G.~J.~Olmo, D.~Rubiera-Garcia and D.~G\'omez S\'aez-Chill\'on,
``Light ring images of double photon spheres in black hole and wormhole spacetimes'',
Phys. Rev. D \textbf{105}, no.8, 084057 (2022),
%doi:10.1103/PhysRevD.105.084057
[arXiv:2202.03809 [gr-qc]].
%26 citations counted in INSPIRE as of 11 Oct 2023

%\cite{Yang:2021cvh}
\bibitem{Yang:2021cvh}
Y.~Yang, D.~Liu, Z.~Xu, Y.~Xing, S.~Wu and Z.~W.~Long,
``Echoes of novel black-bounce spacetimes,''
Phys. Rev. D \textbf{104}, no.10, 104021 (2021)
%doi:10.1103/PhysRevD.104.104021
[arXiv:2107.06554 [gr-qc]].
%6 citations counted in INSPIRE as of 26 Mar 2022

%\cite{Lima:2020auu}
\bibitem{Lima:2020auu}
H.~C.~D.~Lima, C.~L.~Benone and L.~C.~B.~Crispino,
``Scalar absorption: Black holes versus wormholes'',
Phys. Rev. D \textbf{101}, no.12, 124009 (2020),
%doi:10.1103/PhysRevD.101.124009
[arXiv:2006.03967 [gr-qc]].
%13 citations counted in INSPIRE as of 29 Mar 2022

%\cite{Lima:2021las}
\bibitem{Lima:2021las}
H.~C.~D.~Lima, Junior., L.~C.~B.~Crispino, P.~V.~P.~Cunha and C.~A.~R.~Herdeiro,
``Can different black holes cast the same shadow?'',
Phys. Rev. D \textbf{103}, no.8, 084040 (2021),
%doi:10.1103/PhysRevD.103.084040
[arXiv:2102.07034 [gr-qc]].
%35 citations counted in INSPIRE as of 29 Mar 2022

%\cite{Huang:2019arj}
\bibitem{Huang:2019arj}
H.~Huang and J.~Yang,
``Charged Ellis Wormhole and Black Bounce'',
Phys. Rev. D \textbf{100}, no.12, 124063 (2019),
%doi:10.1103/PhysRevD.100.124063
[arXiv:1909.04603 [gr-qc]].
%21 citations counted in INSPIRE as of 26 Mar 2022

%==========================================

%\cite{Guendelman:2015wsv}
\bibitem{Guendelman:2015wsv}
E.~Guendelman, E.~Nissimov, S.~Pacheva and M.~Stoilov,
``Kruskal-Penrose Formalism for Lightlike Thin-Shell Wormholes'',
Springer Proc. Math. Stat. \textbf{191}, 245-259 (2016),
%doi:10.1007/978-981-10-2636-2\_15
[arXiv:1512.08029 [gr-qc]].
%2 citations counted in INSPIRE as of 27 Mar 2022




%\cite{Hoffmann:2017vkf}
\bibitem{Hoffmann:2017vkf}
C.~Hoffmann, T.~Ioannidou, S.~Kahlen, B.~Kleihaus and J.~Kunz,
``Wormholes Immersed in Rotating Matter'',
Phys. Lett. B \textbf{778}, 161-166 (2018),
%doi:10.1016/j.physletb.2018.01.021
[arXiv:1712.02143 [gr-qc]].
%10 citations counted in INSPIRE as of 27 Mar 2022

%\cite{Hoffmann:2018oml}
\bibitem{Hoffmann:2018oml}
C.~Hoffmann, T.~Ioannidou, S.~Kahlen, B.~Kleihaus and J.~Kunz,
``Symmetric and Asymmetric Wormholes Immersed In Rotating Matter'',
Phys. Rev. D \textbf{97}, no.12, 124019 (2018),
%doi:10.1103/PhysRevD.97.124019
[arXiv:1803.11044 [gr-qc]].
%10 citations counted in INSPIRE as of 27 Mar 2022

%\cite{Bakopoulos:2018nui}
\bibitem{Bakopoulos:2018nui}
A.~Bakopoulos, G.~Antoniou and P.~Kanti,
``Novel Black-Hole Solutions in Einstein-Scalar-Gauss-Bonnet Theories with a Cosmological Constant'',
Phys. Rev. D \textbf{99}, no.6, 064003 (2019),
%doi:10.1103/PhysRevD.99.064003
[arXiv:1812.06941 [hep-th]].
%59 citations counted in INSPIRE as of 27 Mar 2022

%\cite{Ibadov:2020ajr}
\bibitem{Ibadov:2020ajr}
R.~Ibadov, B.~Kleihaus, J.~Kunz and S.~Murodov,
``Scalarized nutty wormholes'',
Symmetry \textbf{13}, no.1, 89 (2021),
%doi:10.3390/sym13010089
[arXiv:2012.05178 [gr-qc]].
%3 citations counted in INSPIRE as of 27 Mar 2022

%\cite{Ibadov:2020btp}
\bibitem{Ibadov:2020btp}
R.~Ibadov, B.~Kleihaus, J.~Kunz and S.~Murodov,
``Wormholes in Einstein-scalar-Gauss-Bonnet theories with a scalar self-interaction potential'',
Phys. Rev. D \textbf{102}, no.6, 064010 (2020),
%doi:10.1103/PhysRevD.102.064010
[arXiv:2006.13008 [gr-qc]].
%11 citations counted in INSPIRE as of 27 Mar 2022

%\cite{Anabalon:2018rzq}
\bibitem{Anabalon:2018rzq}
A.~Anabal\'on and J.~Oliva,
``Four-dimensional Traversable Wormholes and Bouncing Cosmologies in Vacuum'',
JHEP \textbf{04}, 106 (2019),
%doi:10.1007/JHEP04(2019)106
[arXiv:1811.03497 [hep-th]].
%20 citations counted in INSPIRE as of 02 Aug 2023



\bibitem{book}
M. Visser, \textit{Lorentzian wormholes: From Einstein to Hawking}, AIP press [now Springer], New York (1995).

%\cite{Boonserm:2018orb}
\bibitem{Boonserm:2018orb}
P.~Boonserm, T.~Ngampitipan, A.~Simpson and M.~Visser,
``Exponential metric represents a traversable wormhole'',
Phys. Rev. D \textbf{98}, no.8, 084048 (2018),
%doi:10.1103/PhysRevD.98.084048
[arXiv:1805.03781 [gr-qc]].
%38 citations counted in INSPIRE as of 02 Dec 2022


%\cite{Morris:1988cz}
\bibitem{Morris:1988cz}
M.~S.~Morris and K.~S.~Thorne,
``Wormholes in space-time and their use for interstellar travel: A tool for teaching general relativity'',
Am. J. Phys. \textbf{56}, 395-412 (1988).
%doi:10.1119/1.15620;
%1765 citations counted in INSPIRE as of 02 Dec 2022

\end{thebibliography}
\end{document}